\documentclass[manuscript, screen]{acmart}

\def\BibTeX{{\rm B\kern-.05em{\sc i\kern-.025em b}\kern-.08emT\kern-.1667em\lower.7ex\hbox{E}\kern-.125emX}}

\setcopyright{acmcopyright}
\acmJournal{TOG}
\acmYear{2019}\acmVolume{0}\acmNumber{0}\acmArticle{0}\acmMonth{0}
\acmDOI{00.0000/0000000.0000000}

\usepackage{todonotes}
\usepackage{upgreek}
\usepackage{amsmath}
\usepackage{algorithm}
\usepackage{algorithmic}
\usepackage{underscore}
\usepackage{ulem}

\usepackage{caption}
\usepackage{subfigure}
\usepackage{graphicx}
\usepackage{notoccite}
\begin{document}

%
% The "title" command has an optional parameter, allowing the author to define a "short title" to be used in page headers.
\title{Low Overhead Online Data Flow Tracking for Intermittently Powered Non-volatile FPGAs}

%
% The "author" command and its associated commands are used to define the authors and their affiliations.
% Of note is the shared affiliation of the first two authors, and the "authornote" and "authornotemark" commands
% used to denote shared contribution to the research.
\author{Xinyi Zhang}
\email{xinyizhang@pitt.edu}
\affiliation{
  \institution{University of Pittsburgh}
  \streetaddress{4200 Fifth Ave}
  \city{Pittsburgh}
  \state{Pennsylvania}
  \postcode{15260}
}

\author{Clay Patterson}
\email{pclaya@ostatemail.okstate.edu}
\affiliation{
  \institution{Oklahoma State University}
  \streetaddress{}
  \city{Stillwater}
  \state{Oklahoma}
  \postcode{74078}
}

\author{Yongpan Liu}
\email{ypliu@tsinghua.edu.cn}
\affiliation{
  \institution{Tsinghua University}
  \streetaddress{Circuits and Systems Division, Electronic Eng. Dept}
  \city{Beijing}
  \state{China}
  \postcode{100084}
}

\author{Chengmo Yang}
\email{chengmo@udel.edu}
\affiliation{
  \institution{University of Delaware}
  \streetaddress{201C Evans Hall}
  \city{Newark}
  \state{DE}
  \postcode{19716}
}

\author{Chun Jason Xue}
\email{ jasonxue@cityu.edu.hk}
\affiliation{
  \institution{City University of Hong Kong}
  \streetaddress{Department of Computer Science, Tat Chee Ave, Kowloon}
  \city{Hong Kong }
  \state{}
  \postcode{}
}

\author{Jingtong Hu}
\email{jthu@pitt.edu}
\affiliation{
  \institution{University of Pittsburgh}
  \streetaddress{4200 Fifth Ave}
  \city{Pittsburgh}
  \state{Pennsylvania}
  \postcode{15260}
}

%
% By default, the full list of authors will be used in the page headers. Often, this list is too long, and will overlap
% other information printed in the page headers. This command allows the author to define a more concise list
% of authors' names for this purpose.
% \renewcommand{\shortauthors}{Xinyi Zhang, et al.}

%
% The abstract is a short summary of the work to be presented in the article.
\begin{abstract}
Energy harvesting is an attractive way to power future IoT devices since it can eliminate the need for battery or power cables. 
However, harvested energy is intrinsically unstable. While FPGAs have been widely adopted in various embedded systems, it is hard to survive unstable power since all the memory components in FPGA are based on volatile SRAMs. 
The emerging non-volatile memory based FPGAs provide promising potentials to keep configuration data on the chip during power outages. 
Few works have considered implementing efficient runtime intermediate data checkpoint on non-volatile FPGAs. 
To realize accumulative computation under intermittent power on FPGA, this paper proposes a low-cost design framework, Data-Flow-Tracking FPGA (DFT-FPGA), which utilizes binary counters to track intermediate data flow. 
Instead of keeping all on-chip intermediate data, DFT-FPGA only targets on necessary data that is labeled by off-line analysis and identified by an online tracking system. 
The evaluation shows that compared with state-of-the-art techniques, DFT-FPGA can realize accumulative computing with less off-line workload and significantly reduce online roll-back time and resource utilization. 

\textit{This paper has been accepted by ACM Journal on Emerging Technologies in Computing Systems (JETC).}
\end{abstract}



%
% Keywords. The author(s) should pick words that accurately describe the work being
% presented. Separate the keywords with commas.

%
% A "teaser" image appears between the author and affiliation information and the body
% of the document, and typically spans the page.
%%  \includegraphics[width=\textwidth]{sampleteaser}
%%  \caption{Seattle Mariners at Spring Training, 2010.}
%%  \Description{Enjoying the baseball game from the third-base seats. Ichiro Suzuki preparing to bat.}

%
% This command processes the author and affiliation and title information and builds
% the first part of the formatted document.
\maketitle

\section{Introduction}
\label{sec:introduction}
FPGAs have been widely adopted in various embedded systems that are powered by batteries. However, in the emerging Internet of Things (IoT)\cite{xu2018scaling,xu2017edge,xu2017efficient}, which is full of tiny, cost-sensitive, and space constrained widgets, batteries are no longer an ideal power supply due to poor scalability, recharging and safety concerns. 
Out of all possible alternatives, energy harvesting systems are becoming one of the most promising candidates because they convert
ambient energy from their surroundings.
Some of the widely known energy harvesting techniques include Photovoltaics (PV), Thermoelectric generators (TEGs), and Piezoelectric (PZ). 
A device equipped with these harvesters can utilize the converted energy directly or recharge its energy storage (e.g. capacitors).
The ease of access to power makes it a very competitive power source for portable devices.
However, there are two primary challenges for such energy harvesting systems: unstable power and low power input.

Even though there are ultra-low power FPGAs such as the Lattice iCE40 series, which can work in $\mu W$~\cite{lattice}, the unpredictability of
available energy renders the power intermittent which will interrupt computations.
The intermittent power will interrupt computations.
In such a condition, long computations may be prohibited since the intermediate data will be lost and the computation has to start over from the beginning.
Thus, it is essential to preserve the FPGA $configuration$ $data$ and $intermediate$ $data$ during a power outage.
$Configuration$ $data$ keeps the functionality of a FPGA chip and $intermediate$ $data$ is the data generated during computation.
By keeping both, long computations can be achieved by retrieving a checkpoint after power resumes.

The non-volatile memory based FPGAs (NV-FPGAs) are natural candidates to address this challenge.
With the substitution of NVMs such as ReRAM, STT-RAM, and PCM for SRAMs, $configuration$ $data$ can be retained locally on the chip with benefits of low leakage power, short critical path, and small area, etc~\cite{Tang2014,Tang2016, zhao2009spin,guo2010resistive, gaillardon2010,Chen2010}.
% ,Gaillardon2010
Therefore, costs associated with loading $configuration$ $data$ from off-chip flash memories are avoided when FPGA recovers from a power outage.
In the existing NV-FPGAs and traditional FPGAs, $intermediate$ $data$ is held by registers which consist of volatile flip-flops (FF).
Like $configuration$ $data$, $intermediate$ $data$ needs to be saved during power is lost or weak in order to resume system state after the power comes back.
To reserve intermediate data,
non-volatile flip-flops (NV-FFs) have been integrated on processors.
Such non-volatile processor freezes all registers data locally on the chip if it is shut off~\cite{lee2017reram}. 
% ~\cite{Chien2016,lee2017reram}.
The success of NV-FFs in processors makes it a good candidate for FPGA flip-flop. However, FPGA's register resource is significantly more than the processor's and FPGA resource utilization varies design to design. 
% Back-up strategy in~\cite{lee2017reram} is too energy consuming.
Freezing all registers on FPGA will waste rare energy in energy harvesting system.

To improve efficiency in preserving $intermediate$ $data$ and reducing the roll-back impact from power interrupt, this paper proposes DFT-FPGA, a data flow tracking methodology on FPGA via High-Level-Synthesis (HLS).
As HLS takes software functions as inputs and compiles it to Register-Transfer-Level (RTL) design.  
DFT-FPGA builds data flow trackers for such functions and a control unit to parse trackers' status in HLS.
With an offline mapping of functions to trackers, DFT-FPGA can online track the $intermediate$ $data$ inside a function via a tracker. Then, instead of all data, only a set of tracked $intermediate$ $data$ in registers will be locally stored in non-volatile flip-flops. In this way, the cost to preserve $intermediate$ $data$ can be significantly reduced. The main contributions of this work are as  follows:

\begin{itemize}
 \item Design of binary counter based tracker framework which tracks data flow in FPGA.
 \item Design of control unit which stores the mapping of intermediate data and its on-chip physical address.
 \item Design of an off-line intermediate data to tracker mapping algorithm.
 \item Design of function split and merge method for DFT-FPGA.
  \item A demonstration of the performance of NV-FF based FPGA.
 \item A demonstration of the efficiency of FPGA design on representative benchmarks.
 \end{itemize}

The rest of the paper is organized as follows. Section~\ref{sec:preliminary} presents FPGA in an energy harvesting system, non-volatile FPGA, High-Level-Synthesis, and the motivation of DFT-FPGA. 
Section~\ref{sec:methods} presents the DFT-FPGA framework. Section~\ref{sec:algorithm} presents the function-to-tracker mapping algorithms and Section~\ref{sec:exp} presents the evaluation results. 

\section{Preliminary}
\label{sec:preliminary}

We will first present FPGA in an energy harvesting system in subsection~\ref{sec:harvesting}. Then works related to NV-FPGA and NV-FF architecture will be presented in subsection~\ref{sec:fpga}. The background of High-Level-Synthesis will be introduced in subsection~\ref{sec:HLSsec} and the motivation of the proposed work will be presented in subsection~\ref{sec:example}.

\subsection{FPGA and Energy Harvesting System}
\label{sec:harvesting}
FPGA is widely adopted in open field applications such as wireless network platforms~\cite{patil2019fpga,bengherbia2017fpga,obeid2016towards}.
In this application scenario, the energy harvesting system outperforms other resources such as battery and cable thanks to its perpetual power supply. 
In energy harvesting systems, ambient energy such as solar, wind, mechanical strain, ambient radiation, and human motion can be harvested to power the energy consumer and its peripheral devices.
Though energy harvesting can provide a perpetual power supply,
when compared to systems powered by cable, energy harvesting system may only harvest and supply a small amount of energy.
Moreover, the passive energy harvesting approach makes the harvested energy unpredictable and unreliable.
It is observed that small harvesters such as a wrist-worn motion harvester can provide about 40 $\mu W$ power with worst-case power outages every $10ms$ in daily activities~\cite{xue2015analysis,ma2017incidental}.
Therefore, a device with the energy harvesting system should not only be able to work in a low-energy mode but also be robust enough under intermittent power.

\begin{figure}[htbp]\small
  \centering
  \includegraphics[width=0.4\columnwidth]{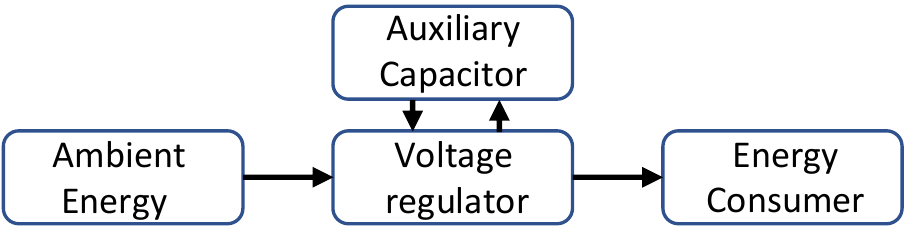}\\
  \caption{An energy harvesting system.  }\label{energyharvesting}
\end{figure}

An energy harvesting system usually consists of an energy source, regulator, capacitor, and energy consumer.
A classical system is shown in Fig.~\ref{energyharvesting}.
The regulator is a bridge between ambient energy, energy consumer, and an auxiliary capacitor.
Energy consumer is powered by the regulator when harvested energy is sufficient. Meanwhile, the capacitor is fully charged and standby for energy spike or power outage/weak.
If the ambient energy becomes weak, the additional energy from capacitor should  sustain the consumer's rest of work or preserve its current state.
As the computation complexity varies from application to application, the energy needed for each application are different.
Therefore, preserving the current states are widely adopted in designs~\cite{ma2015nonvolatile,maeng2017alpaca,pan2017lightweight,balsamo2015hibernus,xie2015fixing,xie2016checkpoint, pan2017lightweight, Mirhoseini,yuan2016cp}.
Moreover, the energy from capacitors is limited as the size of capacitors in small scale energy harvesting system can be as small as $\mu F$ and under $5v$~\cite{pan2017lightweight}.
For energy consumers like microcontroller, ASIC, and FPGA, their dynamic power varies from cycle to cycle. Evaluating and recording the energy from each cycle to the end can be unrealistic as a regular application can be thousands or even dozens of thousands cycles.

\subsubsection{FPGA architecture}
\label{sec:fpga}
\begin{figure}[htbp]\small
  \centering
  \includegraphics[width=0.9\columnwidth]{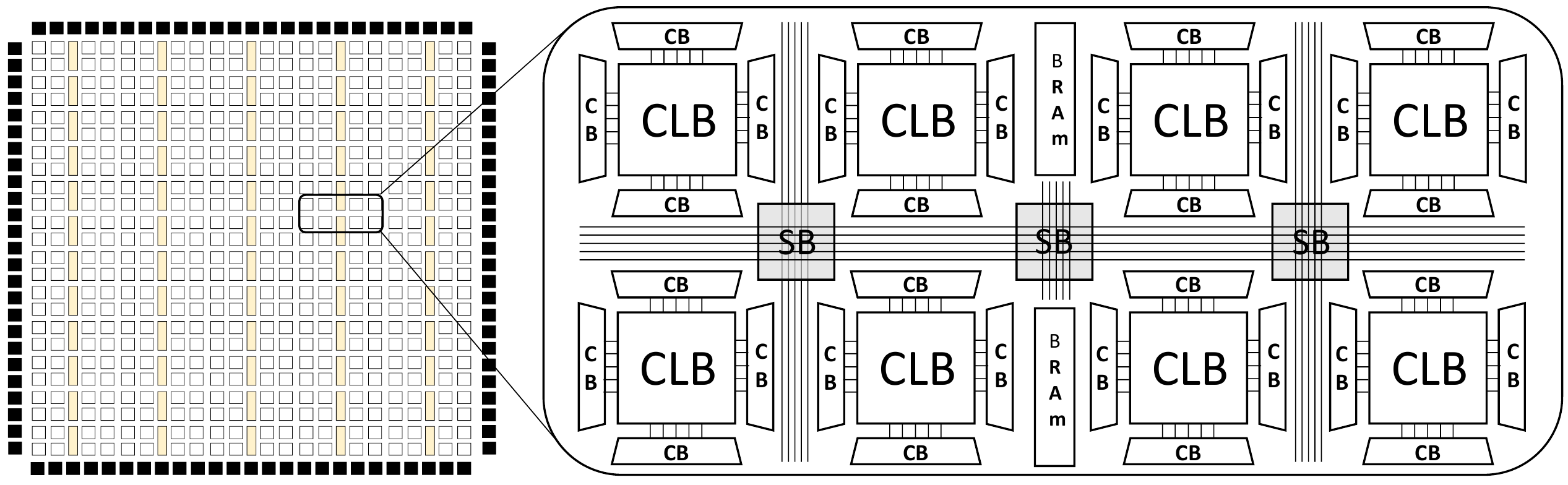}\\
  \caption{An overview of FPGA architecture. }\label{architecture}
\end{figure}

\begin{figure}[htbp]\small
  \centering
  \includegraphics[width=0.90\columnwidth]{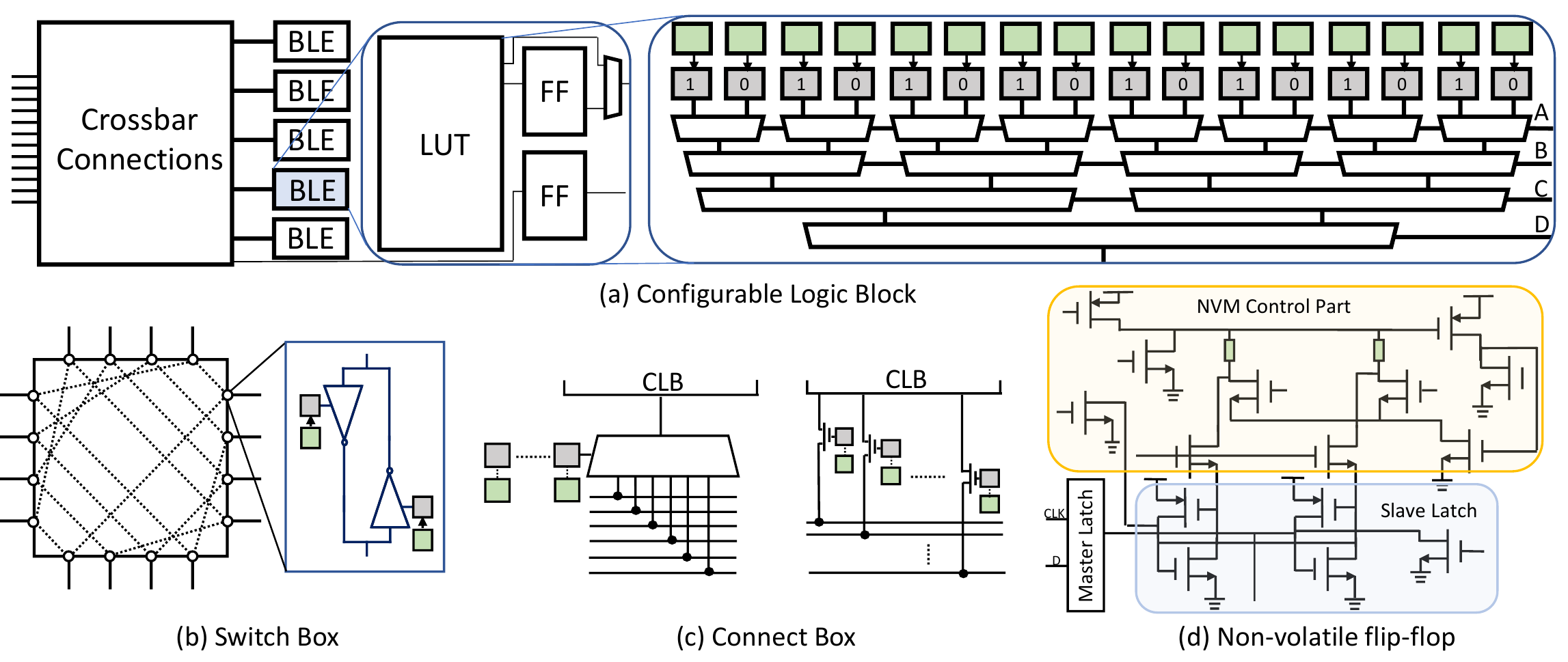}\\
  \caption{FPGA breakdown and Non-volatile. FPGA}\label{nvfpga}
\end{figure}

A basic FPGA architecture consists of Configurable Logic Blocks (CLB), Block RAMs (BRAMs), Connect Boxes (CB), Switch Boxes (SB), and Routing Channels which is shown in Fig.~\ref{architecture}. 
Each CLB is an independent computation group and computation results are transmitted via Routing Channels. 
Switch Boxes redirect the data in Routing Channels to horizontal or vertical channels.
BRAMs in FPGA serve as long-term massive data storage during computing.
In this way, a CLB can reach any other CLB in FPGA.
Computation results can be stored in BRAMs.
The breakdown structure of a CLB from Xilinx FPGA is shown in Figure~\ref{nvfpga} (a).
Look Up Table (LUT) is the smallest programmable computation unit inside a CLB which is wrapped by Basic Logic Element (BLE).
As shown in this 4-input LUT, the LUT can have different logic by changing its mask binary order; different combinations of inputs will lead to different results in logic.
Each LUT is followed by two flip-flops in a BLE which compose registers to support sequential logic in high-frequency computation.
The registers hold the intermediate results from LUTs in each clock cycle.
Then, data is sent out of CLB by Connect Box and this data is routed through Switch Boxes and Routing Channels. 
For a specific application, according to its complexity, 
a number of LUTs, flip-flops, Switch Boxes, and Routing Channels are selected and activated in FPGA to build its corresponding computation logic and data flow path.
(In Xilinx FPGA, a CLB shown in Fig~\ref{nvfpga} (a) is named as SLICE and two SLICEs form a physical CLB)

For traditional FPGA, SRAMs are used to build all these components.
SRAMs are usually fast but volatile.
Thus, the intrinsic flexibility in reconfiguring makes an FPGA easily switch between applications. 
However, configuring data and computation data is vanishing from the chip when power is lost or weak.
In order to keep the data, part of the FPGA components need to be replaced by non-volatile memory.

\subsubsection{Non-volatile FPGA}
The study of non-volatile FPGA has been conducted by Gaillardon, Cong, et, al~\cite{gaillardon2013design, Cong2011,zand2017radiation,jo2016variation,almurib2016design,rajaei2016radiation,ju2018nvm}.
The existing NV-FPGAs preserve all configuration data on the chip.
As the configuration data for LUT, Connect Boxes, and Switch Boxes are stored in SRAMs bit by bit, by replacing configuration data SRAMs with non-volatile memories (NVMs) such ReRAM and STT-RAM, the configuration data can be permanently retained on FPGA.
Though NVMs writing and reading speed is slower than SRAMs, it can preserve its data even power is lost.

Figure.~\ref{nvfpga} shows how the configuration data SRAMs are replaced by NVMs on-chip~\cite{gaillardon2013design}. 
In this figure, a SRAM is represented by a grey block and an NVM is represented by the green block.
The LUT configuration data is stored in its mask which is held by SRAMs.
After replacing the SRAMs in the mask with NVMs, configuration data  for LUT can be permanently preserved on chip after one programming. 
Figure.~\ref{nvfpga} (b) and (c) show how to preserve Connect Boxes and Switch Boxes configuration data with NVMs. 
For Switch Boxes, after configuring all control SRAMs on the connections between horizontal and vertical tracks, data paths inside Routing Channels are formed.
For Connect Boxes, a SRAM determines the on/off of a path between a CLB and a track.
After configuring all needed Switch Boxes and Connect Boxes, multiple data routing paths are formed.
After adopting NVMs, the routing paths in FPGA can also be preserved.
Thus, the configuration data for an application can be preserved on-chip even power is off. 
After programming once, these NVMs can keep all configurations on-chip which avoids duplicate programming for the same program.

SRAM based BRAMs in FPGA can also be directly replaced with non-volatile BRAMs (NV-BRAM). 
BRAMs are usually working as long-term massive storage on-chip.
STT-MRAM based BRAMs (NV-BRAM) are proposed by Ju, et, al~\cite{ju2018nvm}. 
Compared with SRAM BRAMs, NV-BRAMs have higher density and smaller area.
When substituting an SRAM with an NVM cell, six transistors are replaced with one NVM cell. Thus, for Switch Boxes, Connect Boxes, LUTs, and BRAMs, they can have a smaller area and be closer to each other in physical distance.
In this way, the physical data routing distance is reduced, bringing additional benefits such as higher working frequency and more energy efficiency. 
Furthermore, NVMs are compatible with CMOS technology via back-end-of-line technique~\cite{Wong2012}. 
By growing the NVM cells on the top of the chip, it adds non-volatility to an FPGA without increasing chip area and further reduces the area of FPGA components.
An NVM based fully functional FPGA architecture is proposed by Cong, achieving 5.18x area savings, 2.28x speedup and 1.63x power savings compared with regular FPGAs~\cite{Cong2011}.

\subsubsection{Non-volatile flip-flops}
\label{nvff}
For the intermediate data, it is usually held in registers which are consisted of flip-flops. 
Non-volatile flip-flops are proposed by Qazi, Bartling, et, al~\cite{lee2017reram,onkaraiah2012bipolar,Jabeur2014, Chien2016}. 
Contrary to configuration data, intermediate data are frequently refreshed and the data flip frequency can be hundreds of mega hertz (MHz). 
That is, at each clock cycle, the data in a register is refreshed. Therefore, the NVMs writing and reading time may bring delay which hinders the FPGA work frequency. 
It is observed that a 65$nm$ technology based ReRAM NV-FF has achieved writing time as good as 4$\mu$s seconds and 46.2 $pico$ $joul$ per bit~\cite{lee2017reram}. 
Meanwhile, a regular FPGA usually works at hundreds of $MHz$, i.e working under $ns$ clock period. With such performance, if NVMs writing happens in NV-FF every clock cycle, the FPGA frequency may be degraded to dozens of MHz.

The architecture of NV-FF is shown in Fig.~\ref{nvfpga} (d).
Based on a master-slave flip-flop, two pieces of NVMs are integrated to its slave logic.
By adding extra NVM control part, data in slave logic can be optionally stored into or retrieved from NVM cell.
It works like regular flip-flop if NVM writing is not triggered and it can write the data to NVMs or recover data from NVM with a trigger.
When writing to NVM is triggered, the clock will be hung and data is written to NVM and vice versa.
This avoids the writing to NVMs every clock cycle and the working frequency is not influenced if NVM writing is not triggered.
The NVM brings non-volatility but also larger logic area.
It is observed in~\cite{lee2017reram} that a NV-FF based processor has an extra 39\% area for a single FF but less than 10\% extra area for the whole chip.
The 39\% single FF area overhead is brought by the NVM's control logic and the NVM is growing on the top of the chip, which won't bring extra area.
Though NVM writing is not triggered at every clock cycle, the area overhead from the control logic will increase its intrinsic delay and the size of a CLB, leading to bigger FPGA size, which increases the routing distance. This may decrease the FPGA working speed.  
The impacts on FPGAs working frequency after bringing in NV-FF is evaluated in section~\ref{sec:exp}.  

For each NV-FF, it can be triggered to work as regular, store, or retrieve mode. 
If other FPGA components are non-volatile,
by storing the intermediate data to the NVM if power is weak, the chip state can be held on board and be retrieved later.
However, selecting a single flip-flop in an FPGA chip is not accessible as 
a single flip-flop can not be indexed inside a CLB.
Moreover, FPGA's flip-flop resource can number in the thousands and tens of thousands.
In the proposed design, selected flip-flops need to be stored in order to keep intermediate data and flip-flop storing and retrieve are executed SLICE by SLICE.
This is due to FPGA tools like Vivado packing flip-flops into the same SLICE~\cite{floor}. 
A physical CLB in Xilinx FPGA contains two SLICEs and each SLICE can be indexed on the chip. 
In general, 4 LUTs and 8 flip-flops or 8 LUTs and 16 flip-flops are placed in a SLICE.
Therefore, the flip-flop write and retrieve are triggered SLICE by SLICE in the proposed design.

%  As there are usually 8 or 16 flip-flops inside a CLB~\cite{xilixnff}, additional selection logic circuit will further increase the CLB area.

In DFT-FPGA, the SLICE addresses are acquired after FPGA synthesis and pre-loaded to FPGA.
During online intermediate data tracking, such SLICE addresses can be read out after parsing trackers' status. Then, the flip-flops in these SLICEs can be triggered in writing or retrieving.

\subsection{High-Level-Synthesis (HLS)}
\label{sec:HLSsec}
\begin{figure}[htbp]\small
  \centering
  \includegraphics[width=0.95\columnwidth]{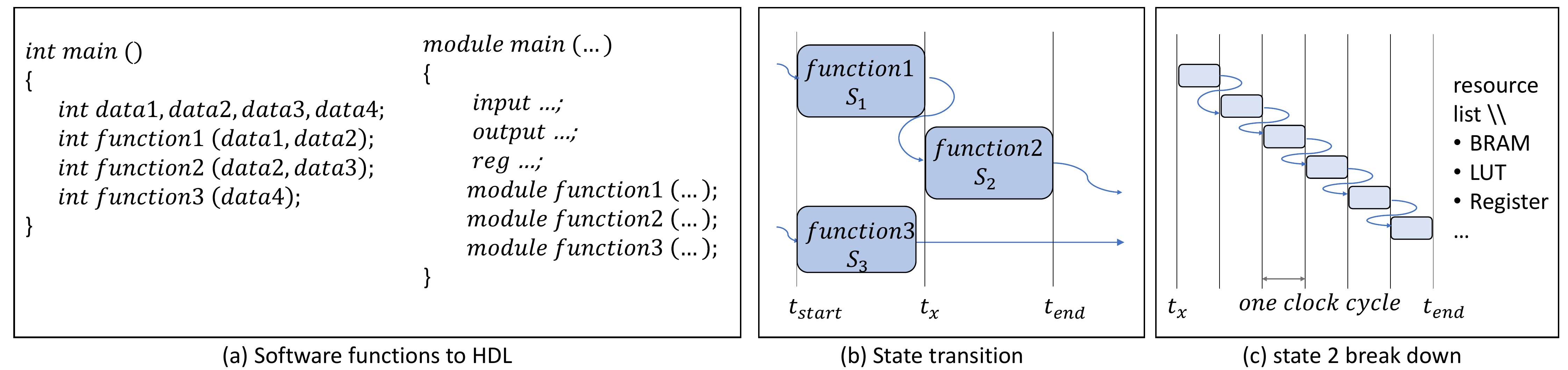}\\
  \caption{High-Level-Synthesis.}\label{hls}
\end{figure}

High-Level-Synthesis converts software language such as C/C++ to Hardware Description Language (HDL) like Verilog and VHDL. HLS's efficiency and accuracy has been verified in modern FPGA applications~\cite{nane2015survey,navarro2013high,jiang2019achieving,jiang2019accuracy}.
It takes software functions ${F}$ as inputs to generate HDL modules $M$ and state transition flow $S$.
After HLS, a program is split into multiple modules according to program hierarchy.
These modules are interpreted as different states in the state transition control.
State transition control initializes and terminates the modules during FPGA operation.
In HLS, each function under top function is generally compiled to a standalone module. 
Based on the data dependency between modules (functions), these modules are triggered in parallel or sequential in state transition control.
A basic HLS FPGA design with its data flow control is shown in Fig.~\ref{hls}~\cite{Canis2013}.
% $f_{\textrm{pre}}(x)$

In this program, there are three functions $F_1$, $F_2$ and $F_3$ under top function $F_{main}$.
After HLS, functions are converted to modules as shown in Fig.~\ref{hls} (a).
In the top function, $F_1$ and $F_2$ have data dependency and $F_3$ is independent of $F_1$ and $F_2$. 
Therefore, as shown in Fig.~\ref{hls} (b) state transition control
, $F_2$ has to be placed after $F_1$ in state flow $S$. Meanwhile, $F_3$ can start with $F_1$. 
In states $S_1$, $S_2$ and $S_3$, component information in module $M_1$, $M_2$ and $M_3$ such as module names, registers, connections, and cycles can be collected.
An unroll of $S_2$ is shown in Figure.~\ref{hls} (c).
For HLS tools like Vivado/LegUp, it maps module components to different clock cycles and indicates the data flow among clock cycles.
In the real case, a state like $f_2$ may contain sub-state, each sub-state can be unrolled according to clock cycles.

After High-Level-Synthesis, synthesis turns HDL into the implementation of logical gates, BRAMs, and registers.
It also indicates the physical placement of all modules components.
After implementation, these components are mapped to different SLICEs.
The address of these SLICEs is generally represented as 2-dimension $X_{xx}$$Y_{yy}$ ~\cite{floor}.

The proposed design utilizes state transition control after HLS and module components placement after synthesis.
By analyzing the state transition, DFT-FPGA builds cycle accurate data tracker for each software function.
After synthesis, SLICE addresses related to different registers are acquired.
By online reading the status of trackers and parsing it to SLICE address, the SLICEs holding the intermediate data can be selected and saved.

\begin{figure}[ht]
  \centering
  \includegraphics[width=0.85\columnwidth]{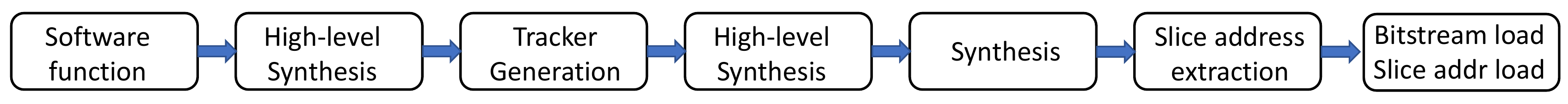}\\
  \caption{DFT-FPGA work flow.}\label{workflow}
\end{figure}

The work flow of DFT-FPGA is shown in the Figure.~\ref{workflow}.
HLS first generates the state transition and state components information. 
Tracker generation builds tracker framework accordingly.
After tracker generation, the source function combined with tracker design are HLSed one more time.
Then, $synthesis$ generates the executable bitstream and provides all SLICE address $X_{xx}$$Y_{yy}$. 
After programming FPGA, DFT-FPGA pre-loads the SLICE addresses to FPGA and starts the computation.
The DFT-FPGA framework will be presented in Section~\ref{sec:methods} and the off-line analysis algorithm will be presented in Section~\ref{sec:algorithm}.

\subsection{Motivation}
\label{sec:example}
\begin{figure}[htbp]
  \centering
  \includegraphics[width=1\columnwidth]{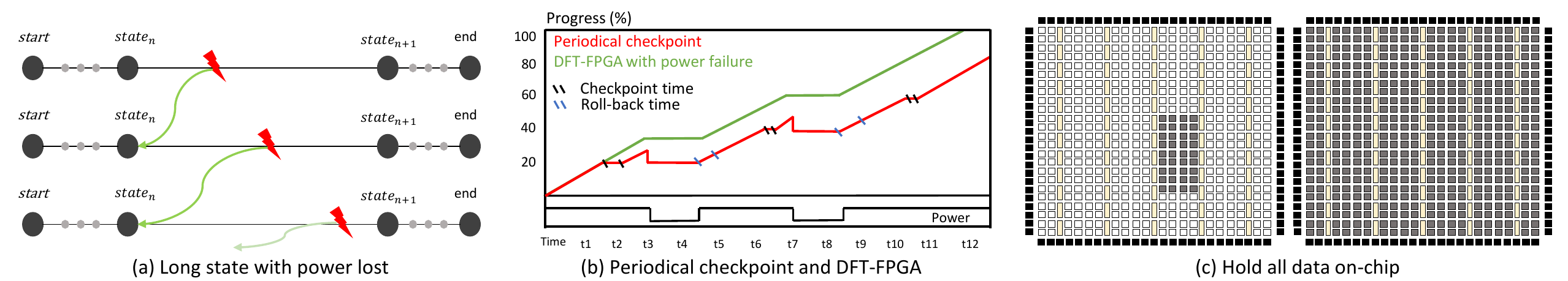}\\
  \caption{The motivation example.}\label{motivate}
\end{figure}

Existing designs preserve chip state by placing checkpoints in computation~\cite{ma2015nonvolatile,maeng2017alpaca,pan2017lightweight,xie2016checkpoint,ahmed2018towards,Mirhoseini,yuan2016cp}.
Azalia et al. propose $Chime$, which places checkpoints in Register-Transfer-Level (RTL) design. 
$Chime$ off-line analyzes the forward computing cost and back-up overhead to determine checkpoint placement. 
By placing checkpoint in different locations, the computation can recover from a near checkpoint if a power outage happens. 
Yuan et al. propose $CP$$-$$FPGA$, a framework that based on HLS result. It places the different states of the main function in different BRAM-based (block RAM) physical areas by modifying FPGA synthesis algorithm.
Each state result is stored in its dedicated BRAM and when power is lost in the mid of a state, the computation can start from its previous state.
In the existing works, the intermediate data back-up is triggered at every checkpoint. 
After retrieving from a checkpoint, the computation can continue. 
While the prior strategies can successfully preserve the intermediate data, long roll-back can easily occur if power is lost in a long state.
This case is shown in Fig.~\ref{motivate} (a), a dot represents the end of a state. 
Checkpoints are placed at each dot. 
For applications with a long state, such strategies may be hard to complete the computation if a power outage happens in the middle of a long state.
Roll-back may occur several times inside a long state until power is stable.
Moreover, by tying a state and a BRAM together, BRAM resources can be insufficient for a large program with multiple states, and periodical data back-up happens even if power is stable.

Fig~\ref{motivate} (b) shows the performance comparison between proposed design and periodical checkpoint design; the red line shows checkpoint technology and the green line shows DFT-FPGA. Power outage happens at $t_3$ and $t_7$. Checkpoints are placed at progress $20\%$, $40\%$ and $60\%$.
As checkpoints are pre-defined, data writing occurs at each checkpoint regardless of the power condition.
The computation starts from its previous checkpoint after power loss.
However, in DFT-FPGA, the checkpoint is avoided and data flow is always tracked. 
Thus, data back-up happens only if power is lost and
computation starts from the progress where power breaks.
Compared with DFT-FPGA, the existing checkpoint placement technologies for FPGA in energy harvesting system suffer from long roll-back time, increasing the complexity of original program, and modifying the FPGA synthesis algorithm.

Preserving the data for the entire chip may also be an option~\cite{lee2017reram}.
By triggering writing and retrieving of flip-flops in all FPGA SLICEs, all the data on chip can be fully reserved.
However, such strategy can be heavily resource intensive.
As FPGA resource utilization varies from design to design, preserving the entire chip causes unnecessary NVM cell writing in unused memories, which will waste rare harvested energy and slow down the writing speed.
As it is shown in Figure.~\ref{motivate} (c), this computation only occupies 15\% of the chip resource, the remaining 85\% components are still stored and retrieved if preserving the whole chip data.
After applying DFT-FPGA, unnecessary back-up can be avoided. 
The benefits and impact after applying DFT-FPGA will be presented in section~\ref{sec:exp}.

Our previous work FC-FPGA applies shift-register like data flow tracker to locate and retrieve intermediate data in RTL level~\cite{zhang2018low}.
In this work, binary counters are adopted to further reduce resource utilization and a full High-Level-Synthesis based work flow is proposed, which significantly reduces the workload.

\section{DFT-FPGA Framework}
\label{sec:methods}

The proposed DFT-FPGA methodology includes both framework and data-to-tracker mapping algorithm designs.
In this section, we will introduce the framework of DFT-FPGA.
The algorithms that work with the framework will be introduced in Section~\ref{sec:algorithm}.

\subsection{Hardware Architecture Overview}

\begin{figure}[ht]
  \centering
  \includegraphics[width=0.9\columnwidth]{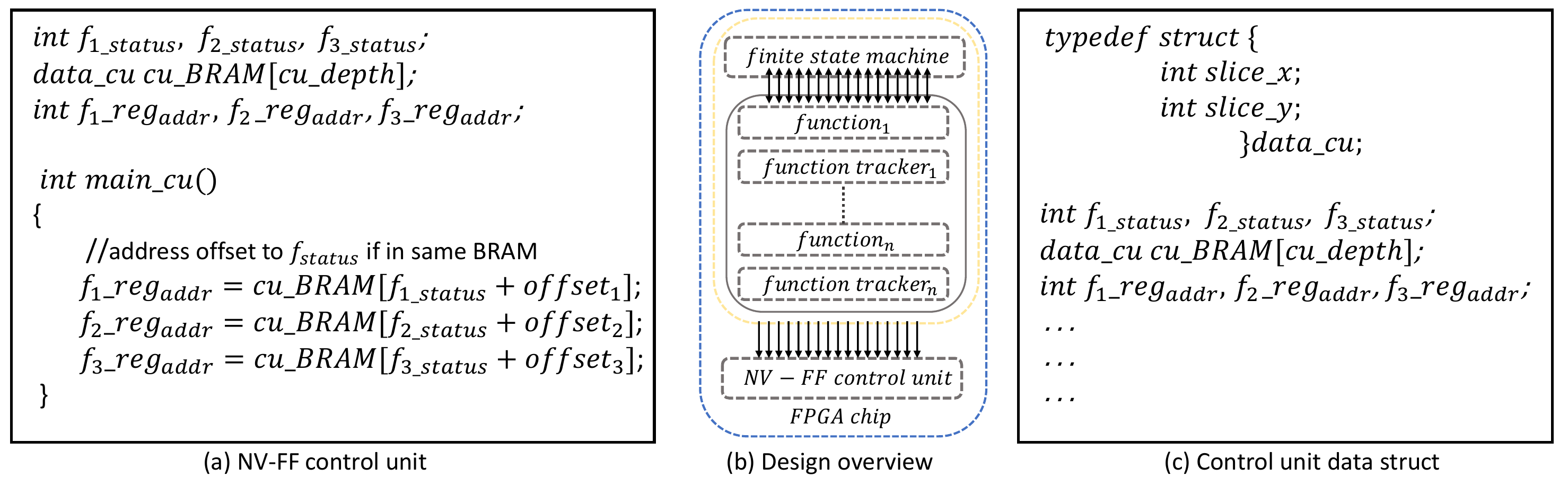}\\
  \caption{Control Unit and DFT-FPGA overview.}\label{cu}
\end{figure}

The proposed design includes function trackers $f$ and NV-FF control unit $CU$ which is shown in Figure~\ref{cu} (b).
In this figure, a finite state machine is generated by HLS in the back-end to control data transition in functions, i.e. state transition between states.
Each function is assigned a function tracker.
Trackers are read by the control unit which pre-loads SLICE addresses and maps them to associated registers.
In DFT-FPGA, at every power outage, the control unit reads function tracker's status and then select the corresponding SLICE to trigger action.
In the proposed design, the control path to NV-FF control part is considered already embedded in NV-FPGA.

\subsection{Function Trackers}
\label{trackers}
As all intermediate data is held by registers, function trackers are designed to track the active registers at each clock cycle.
Therefore, a tracker is built to have the same clock cycles with its corresponding function.
By reading tracker's status, the data flow location in its function can be acquired.

\begin{figure}[ht]
  \centering
  \includegraphics[width=0.9\columnwidth]{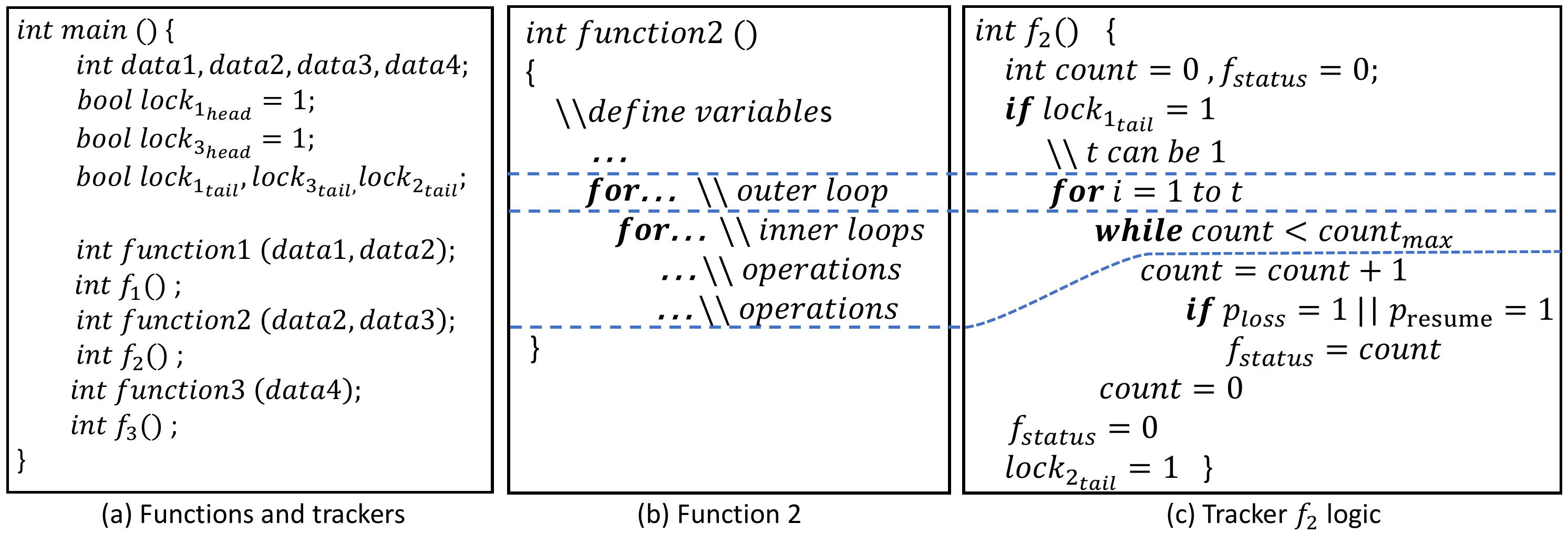}\\
  \caption{Tracker design and asignment in HLS}\label{tracker}
\end{figure}

In DFT-FPGA, each function is assigned a function tracker to trace its data flow.
The tracker is activated simultaneously with its corresponding function and they are terminated at the same time.
The method to insert trackers to a software program is shown in Figure.~\ref{tracker} (a).
Functions $F_1$, $F_2$, and $F_3$ under $F_{main}$ are assigned private trackers $f_1$, $f_2$, and $f_3$. The state transition and timing among $F_{1-3}$ is as it is illustrated in Fig.~\ref{hls} (b).
Therefore, the initialization of trackers $f_{1-3}$ should also follow such orders.
In the proposed design, $lock$ is utilized between trackers to keep trackers initialized in the right order.
The $lock$ in a tracker consists $lock_{head}$ and $lock_{tail}$.
If a tracker is initialized after its anterior tracker, its $lock_{head}$ is the $lock_{tail}$ of the anterior tracker.  
As shown in Fig.~\ref{tracker} (c), 
tracker ${f_2}'s$ $lock_{2_{head}}$ is ${f_1}'s$ $lock_{1_{tail}}$.
After tracker $f_1$ is terminated, $lock_{1_{tail}}$ is set to be 1.
Thus, tracker $f_2$ is always blocked if $F_1$ and $f_1$ are not finished.
As there is no data dependency between function $F_1$ and tracker $f_1$, $f_1$ can start with function $F_1$ simultaneously. 
For trackers which start with the beginning function, such as $f_1$, its $lock_{1_{head}}$ is pre-defined to be 1 to unblock itself.
In this way, all trackers can be initialized and terminated with their corresponding functions.

A function tracker consists of loop arbitration $t$, binary counter $count$, tracker status register $f_{status}$, and tracker lock $lock_{head}$ $lock_{tail}$.
Without loss of generality, trackers can track function with loops or with regular operations.
An example of a function and its tracker logic is shown in Fig.~\ref{tracker} (b) and (c).
This is a function with an outer loop and an inner loop.
In the tracker, the first $loop$ $t$ handles functions with loops. 
Loop iterations $t$ corresponds to the outer $loop$ iteration number in a function.
$Count_{max}$ is the length of all operations under function's outer loop. 
The binary counter increments to $count_{max}$ and then be reset to zero for the next outer $loop$ iteration.
Thus, the binary counter can be reused in all outer loop iterations.
If a function does not contain any loop, $t$ is set to one and $count_{max}$ is the length of the function.
When power is lost during computing, the energy harvesting system sends ${P_{loss}}=1$ to DFT-FPGA.
With ${P_{loss|resume}}=1$, the tracker sends out its status $count$ as $f_{status}$. 
Trackers' $f_{status}$ will be further parsed by control unit.
When reaching the end of tracking, the $f_{status}$ is reset to zero.
In this way, the tracker can count with the function process and send out the process stage when power is loss.
After power resumes $P_{resume}$, trackers data is recovered by NV-FPGA first and the resumed trackers' status are utilized to wake up the stored intermediate data.   

To further reduce the binary size, the binary counter can be defined to lower bit-width such as $4bit$, $8bit$, and $16bit$ according to the length of a function. In most cases, the binary counter size is found to be small as 4-8 $bit$ (tracking length ranges 225-65025 cycles).
The resource utilization, tracking length, and performance of trackers in different sizes will be discussed in section~\ref{sec:exp}.
% When power loss happens, the data flow is freezing as the clock is hanged. 
% By reading trackers' status, control unit can know the status of each function.

\subsection{NV-FF Control Unit} 
The control unit read all trackers' status $f_{status}$.
As trackers' status is cycle accurate, by keeping the active register SLICE address for each clock cycle in control unit, DFT-FPGA can acquire the physical location of intermediate data flow.  
Such address can then be utilized by NV-FPGA to have SLICE registers data stored of retrieved.

The control unit also keeps the mapping relations of function trackers status and registers' SLICE address.
It consists of one or multiple BRAMs and address look-up logic as shown in Figure.~\ref{cu} (a).
All associated SLICE addresses for trackers are pre-loaded to $cu_{BRAM}$ with index $f_{status}$+$offset$.
$Offset$ works as index boundary in $cu_{BRAM}$ when a new tracker is generated. And each $cu_{BRAM}[offset]$ is pre-set to zero. 
Thus, each tracker can have a certain range of storage in $cu_{BRAM}$.
In this figure, all trackers $f_{status}$ is combined with offset including the first tracker $f_1$.
This is because trackers themselves in DFT-FPGA are needed to be reserved as well, $cu_{BRAM}$ keeps trackers SLICE address with index zero to $offset_1$.
The SLICEs where trackers are placed are always stored when power is lost.
By online reading $f_{status}$, corresponding SLICE addresses are parsed by the control unit.
SLICE storing or retrieving can then be operated on these SLICEs.
As tracker $f_{status}$ is zero if it is not triggered or terminated, no action to SLICEs will be executed for such trackers and the control unit will parse zero addresses. 

In DFT-FPGA, one BRAM $cu_{BRAM}$ is instantiated to keep all trackers' mapped SLICE addresses.
During HLS, if the depth of the $cu_{BRAM}$ is too big to be placed in a single physical BRAM, the synthesis tool will automatically expand it to multi-BRAM.
As the SLICE address in FPGA is organized as $X_{xx}$ $Y_{yy}$, a data structure is used in DFT-FPGA to keep address in $X$ and $Y$ direction which is shown in Figure.~\ref{cu} (c).
The resource utilization and performance of the control unit will be discussed in section~\ref{sec:exp}.

After assigning trackers to functions, there are two tasks that need to be accomplished before DFT-FPGA can work properly.
First of all, for a given program with multiple functions, we need to identify which registers will keep intermediate data at a certain clock cycle in each function.
Second, as tracker assignment is arranged according to function hierarchy, the main function should only contain the substantiation of functions. In this task, function merge and function split will be discussed. 

\section{DFT-FPGA off-line analysis}
\label{sec:algorithm}
In previous sections, we show how to generate and assign trackers to functions.
In this section, we will present how to establish mappings between a function and its tracker, and how to merge or split functions in a program.

\subsection{Function to Tracker Mapping}
\label{sec:mapping}
\begin{figure}[ht]
  \centering
  \includegraphics[width=0.8\columnwidth]{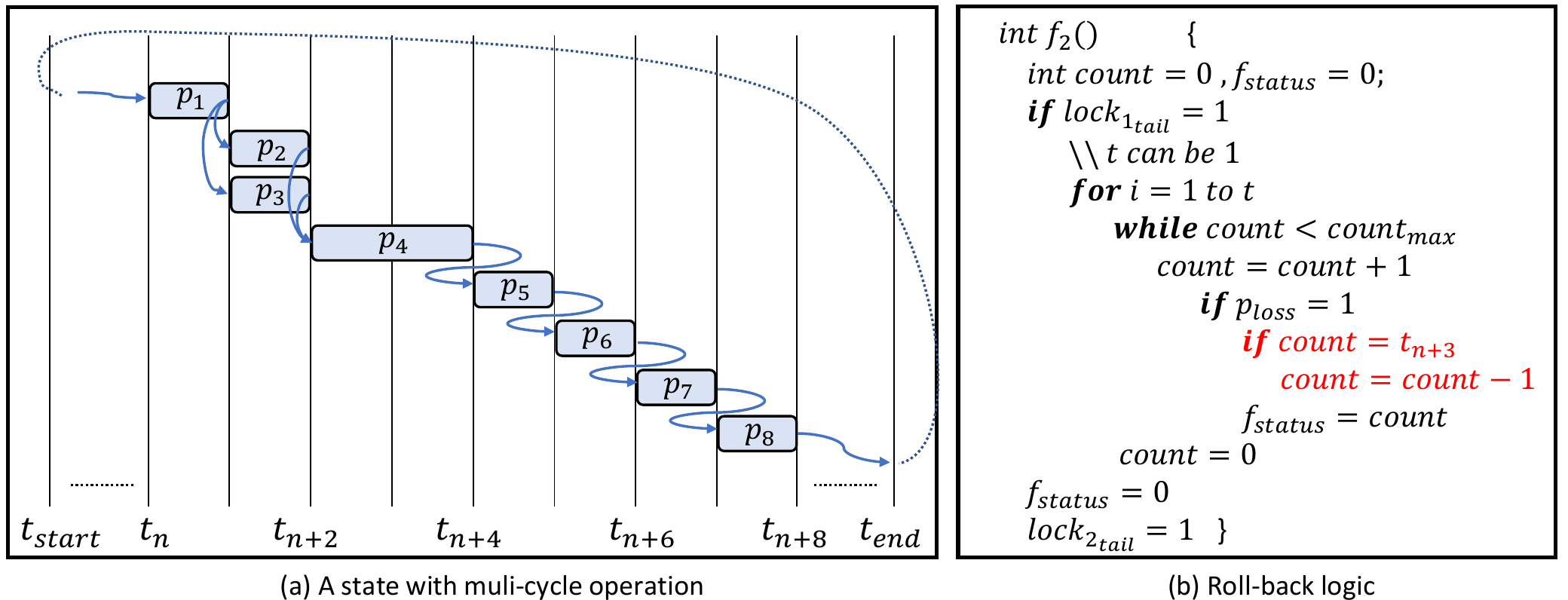}\\
  \caption{A state breakdown and roll-back logic.}\label{algorithm1}
\end{figure}

After analyzing the program hierarchy and the state transition, the activation of trackers are determined.
The mapping of tracker status to data flow in a function can be determined by unrolling a function's state.
A state breakdown is shown in Figure.~\ref{algorithm1} (a).
In this figure, rectangle blocks represent operations inside the function and they are arranged to clock cycles from $t_{start}$ to $t_{end}$. At each clock cycle $t_n$, operations $p_{t_n}$ with its register $reg_{t_n}$ are placed. The connecting arrows indicate the data flow inside a state.
In this figure, $p_2$ and $p_3$ are operations for $p_{t_{n+2}}$; $reg_2$ and $reg_3$ are registers for $reg_{t_{n+2}}$.
Every operation is followed by its register to get its data held at every clock cycle.
For registers $reg_{t_n}$ in a state, they keep the intermediate data within a function at different clock cycles.
Those registers are the target to be tracked by trackers defined with $checkpoint_{t_n}$.
As different registers are triggered during the function process, registers that hold intermediate data for each clock cycle should be determined and those registers' SLICE addresses are stored by the control unit.

In FPGA design, operations can execute in parallel like $p_2$ and $p_3$.
And operations may have multi-cycle length such as $p_4$. 
Therefore, at certain clock cycle, there can be multiple registers $reg_{t_n}$ or previous register $reg_{t_{n^\prime}}$ to hold the intermediate data.
The method to determine $checkpoint_{t_n}$ is illustrated in Algorithm~\ref{alg:checkpoint}.
At each cycle $t_n$, its register $reg_{t_n}$ is added to $checkpoint_{t_n}$ because its operations end at this cycle.
If a multi-cycle operation is cross $t_n$ which starts at $t_{n^\prime}$ and ends at $t_{n^*}$, the $reg_{t_{n^\prime}}$ ahead of this operation is also added to $checkpoint_{t_n}$, e.g,
$checkpoint_{t_{n+2}}$=$reg_2$ $\cup$ $reg_3$;
$checkpoint_{t_{n+3}}$=$reg_2$ $\cup$ $reg_3$.
In this way, the registers holding intermediate data for each clock cycle are acquired.
For multi-cycle operation like $p_4$ at $t_{n+3}$, DFT-FPGA inserts roll-back logic to ensure the consistency between tracker and function after retrieve.
The roll-back logic is shown in Fig.~\ref{algorithm1} (b).
After applying Algorithm~\ref{alg:checkpoint} to all functions and trackers, the mappings between registers to tracker status is established.

If this is a loop function as indicated by the dashed line in Fig.~\ref{algorithm1} (a), $t_{start}$ to $t_{end}$ is tracked by $count$ and its loop iteration is controlled by $t$ in tracker.
The same binary counter $count$ will be called $t$ times.
In this way, DFT-FPGA can scale down the binary counter size and save more flip-flop resources.
By analyzing one iteration's mapping relation, the mapping for the whole function is acquired.
During synthesis, these registers' SLICE address can be acquired.
After that, the mappings between trackers status and SLICE address can be established. 
Then, applying the mapping algorithm, registers are mapped to tracker.
However, for registers barren computation, 
DFT-FPGA can choose to not assign tracker to if tracker flip-flop resources are more than target registers.
Such cases are studied in section~\ref{sec:exp}.

\begin{algorithm}
\caption{$Checkpoint_t$ determination}\small
\label{alg:checkpoint}
\begin{algorithmic}
\STATE {Input: function state $S$, start point $t_{start}$, end point $t_{end}$, operations $p_{t_n}$, registers $reg_{t_n}$ 
\\Output: $checkpoint_t$
\\Define: $start<n<end$, $start<n^\prime<n$, $n<n^*<end$}

\FOR  { $t_n$ $\in$ $S$}
    \STATE {$checkpoint_t$.append($reg_{t_n}$)}
    \FOR {$t_{n^\prime}$ $\in$ $S$}
        \FOR {$t_{n^*}$ $\in$ $S$}
        \IF {$reg_{t_{n^\prime}}\diamondsuit p_{t^*}$ = 1}
        \STATE {$checkpoint_t$.append($reg_{t_{n^\prime}}$)}
         $ \verb|\| \verb|\| \diamondsuit $ 

        % \COMMENT{\diamondsuit represent data path} 
        \ENDIF
        \ENDFOR
    \ENDFOR
\ENDFOR
\RETURN{$checkpoint_t$}

\end{algorithmic}
\end{algorithm}

After analyzing the mapping between a tracker and its function, the trackers for the main function $F_{main}$ need to be arranged.
The proposed DFT-FPGA will tune the hierarchy of $F_{main}$ to make it suitable for tracker assignment. The next task includes function split and function forming.

\subsection{Function Split and Merge}
\label{sec:secsplit}
\subsubsection{Function Split}
\begin{figure}[ht]
  \centering
  \includegraphics[width=0.9\columnwidth]{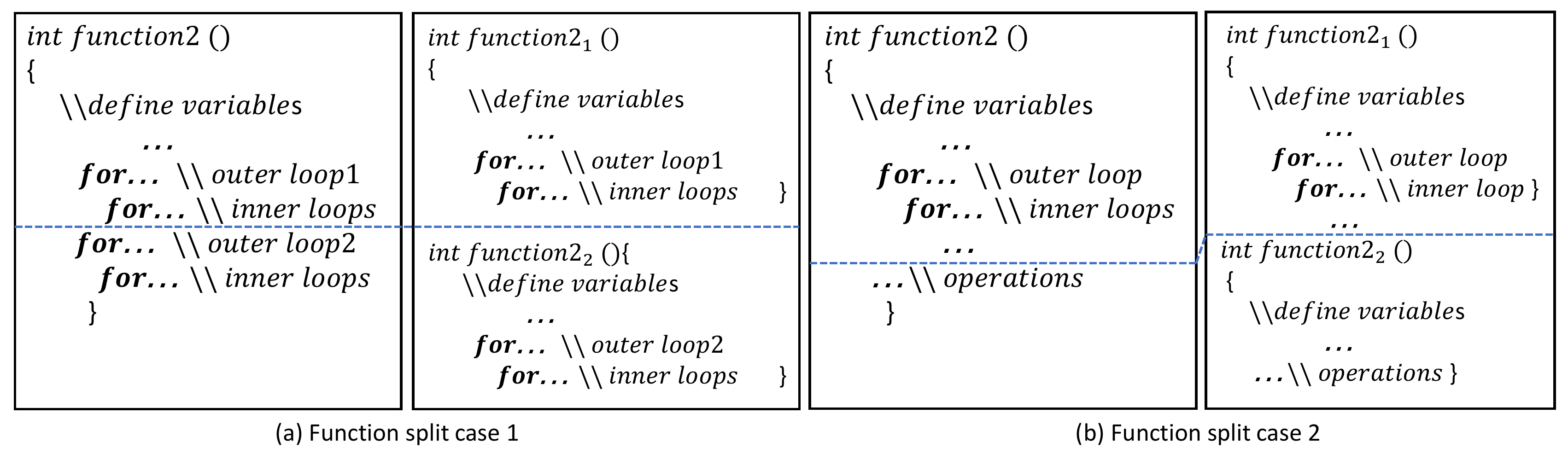}\\
  \caption{Function split.}\label{split}
\end{figure}

As software language is normally flexible in hierarchy and coding style.
A program's hierarchy may need to be tuned and then DFT-FPGA can be applied.
A function needs to be split if there are independent loops or operations aside of a loop. Such a case is shown in Figure.~\ref{split} (a) and (b).
For independent loops as shown in Fig.~\ref{split} (a), the function is split according to outer loops.
As the tracker in DFT-FPGA is designed to track one outer loop with its all inner operations, a function is split to multiple functions according to the number of independent loops. 
Then, trackers will be assigned to split functions.
If a function contains more than loop functions, the function needs to be split based on the boundaries between loop and other operations.
A case where a loop is followed by other operations is shown in Fig.~\ref{split} (b). It needs to be split into two functions. 
Through splitting functions, the proposed tracker logic can be successfully applied.
\subsubsection{Function Merge}
\begin{figure}[ht]
  \centering
  \includegraphics[width=0.6\columnwidth]{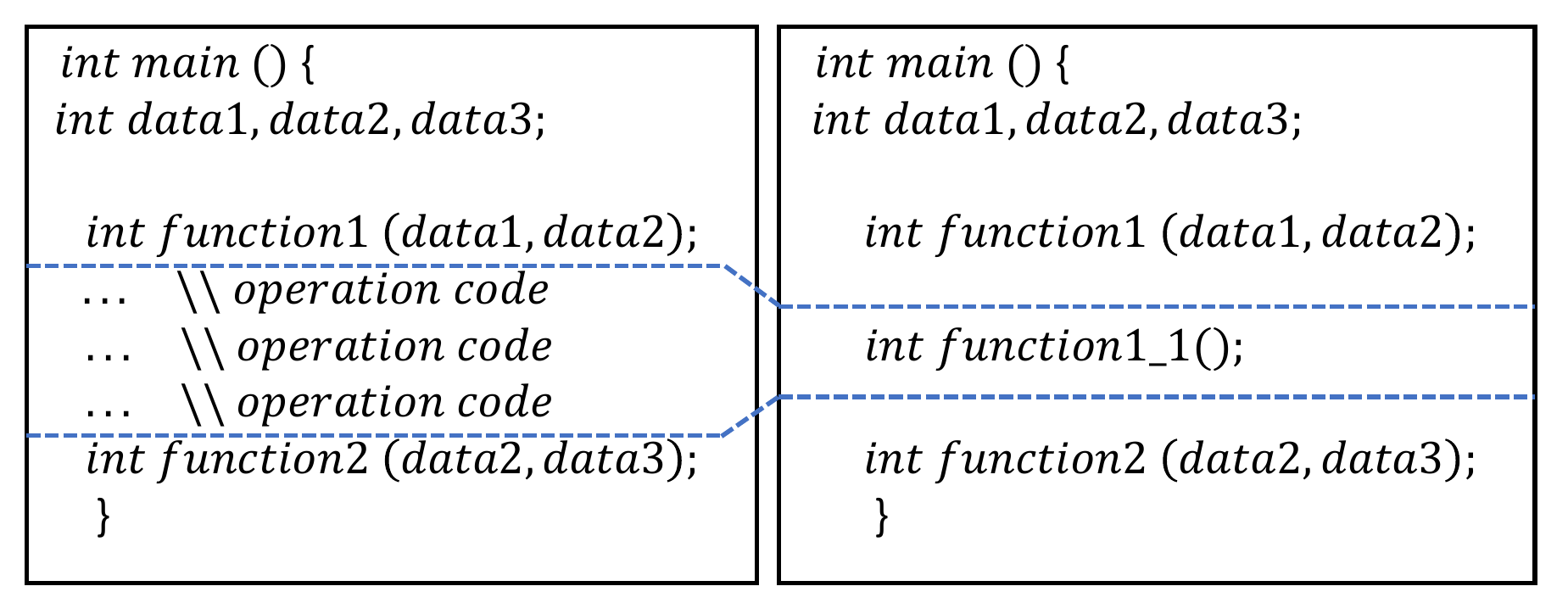}\\
  \caption{Function merge.}\label{merge}
\end{figure}

Under the main function, there can be operations between functions.
As the state transition is arranged between functions, those non-function operations will be automatically merged into a function by HLS.
This will cause obfuscation in function-to-tracker mapping and the auto-merged function cannot be directly applied with trackers.
Such a case is shown in Fig~\ref{merge}.
By wrapping these operations to a function, state transition can be arranged in different functions.
Then, trackers can be built and assigned to all tuned functions.

\section{Experiment}
\label{sec:exp}
In this section, we first evaluate the performance of FPGA architecture with NV-FF in subsection~\ref{nvfpgaperformance}. 
Second, we evaluate the proposed tracker and control unit resource utilization and performance in subsection~\ref{trackerperformance}. 
Third, we evaluate the resource utilization and performance after applying the proposed DFT-FPGA on different benchmarks in subsection~\ref{dftfpgaperformance}. 
Last we evaluate the performance comparison between the periodical checkpoint and the proposed design in subsection~\ref{compare}. 
In experiments, tool Verilog-To-Routing (VTR) is used in evaluating the FPGA architecture with NV-FF in subsection~\ref{nvfpgaperformance}.  
Vivado HLS is used in remaining evaluations for analyzing and generating DFT-FPGA and analyzing benchmarks. 
Vivado is used in FPGA synthesis and getting all registers SLICE address.

\subsection{NV-FF FPGA performance}
\label{nvfpgaperformance}
\begin{figure}[htbp]
  \centering
  \includegraphics[width=1\columnwidth]{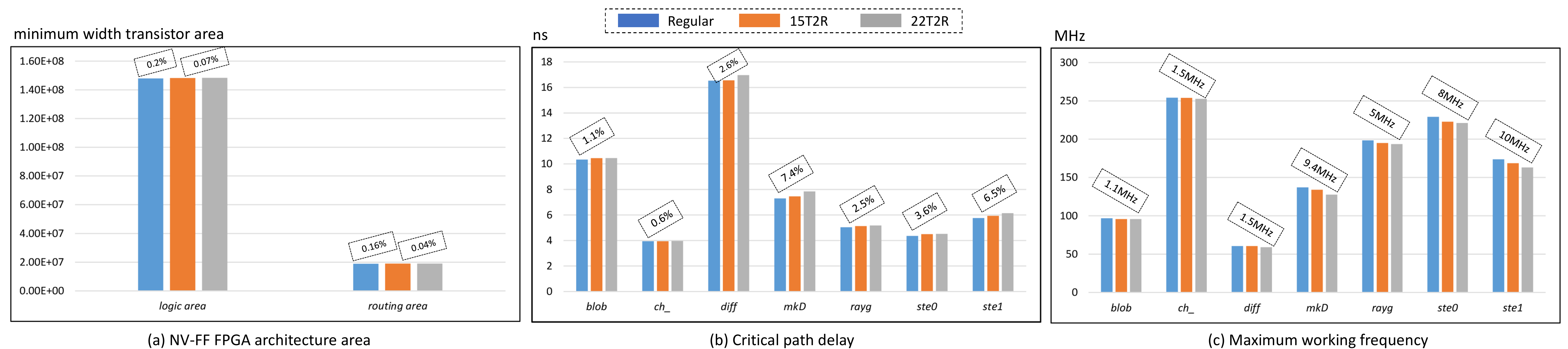}\\
  \caption{NV-FPGA performance.}\label{expa}
\end{figure}

In this subsection, we base on FPGA architecture $K6-frac-N10-mem32K-40nm$ from VTR, by adding delay to its D flip-flop module and increasing CLB area to evaluate the impact from NV-FFs.
As presented in section~\ref{nvff}, the control part in NV-FF brings extra area.
Increasing the area of a single flip-flop can degrade its timing performance but flip-flop area size is not simulated in VTR.
Therefore, we add extra delay to each single flip-flop according to the area scaling up ratio $39\%$ and $49\%$~\cite{lee2017reram,49dff}.
The extra $39\%$ case is caused by additional $15T2R$, which is 15 transistors and 2 ReRAMs~\cite{lee2017reram}.
The extra $49\%$ case is caused by additional $22T2R$, which  is 22 transistors and 2 ReRAMs~\cite{49dff}.
The increase of flip-flop size also brings a larger CLB area, which will lead to longer routing distancedistance and may degrade FPGA's working frequency.
In the evaluated FPGA architecture, each CLB contains 10 flip-flops. For the two types of NV-FF structure, due to VTR adopts $minimum$ $width$ $transistor$ $area$~\cite{khan2017study} to define the size of components, we increase the CLB size by $15$ $minimum$ $width$ $transistor$ $area$ and $22$ $minimum$ $width$ $transistor$ $area$ accordingly.
The FPGA architecture size increment is shown in Figure.~\ref{expa} (a). In this figure, the blue column represents the base architecture, the orange column represents the $15T2R$ architecture, and the grey column represents the $22T2R$ architecture. We observe a $0.2\%$ logic area increasing from base to $15T2R$ and $0.07\%$ from $15T2R$ to $22T2R$. Routing area increases $0.16\%$ from base to $15T2R$ and $0.04\%$ from $15T2R$ to $22T2R$.

To evaluate the impact on FPGA working frequency which suffers from longer flip-flop delay and routing distance,
we apply seven benchmarks in VTR on three FPGA architectures, respectively.
The critical path delay and maximum working frequency are shown in Fig.~\ref{expa} (b) and (c). In the evaluated benchmarks, we observe that $15T2R$ architecture causes less than $3\%$ of additional critical path delay and $22T2R$ architecture causes less than $7.4\%$ additional critical path delay compared to base architecture.
As the critical path delay determined the maximum working frequency, we also show the maximum working frequency in Fig.~\ref{expa} (c). We can observe that less than $6.5MHz$ degradation is caused by $15T2R$ and less than $10MHz$ is caused by $22T2R$.
From the evaluation, we can see that integrating NV-FF on FPGA brings a bit performance degradation, i.e, several $MHz$. However, as most designs running on FPGA work at hundreds of $MHz$~\cite{nane2015survey, jiang2019achieving, cadambi2010programmable,chakradhar2010dynamically}, such degradation has little impact on overall performance and achieves non-volatility of flip-flops.

\subsection{Tracker and Control Unit Evaluation}
\label{trackerperformance}
In the previous section, we evaluated the performance of NV-FF based FPGA architecture.
In this section, we will evaluate the resource utilization and timing performance of DFT-FPGA framework.
The evaluation is based on FPGA chip $xc7z020clg484$.
Table~\ref{Trackerresource} shows the flip-flop, LUTs usages, and maximum tracking cycles  when a tracker size scales from $4bit$ to $9bit$. The resource of $xc7z020clg484$ is also listed.
In the evaluation, we observe
less than $0.03\%$ of flip-flops and less than $0.21\%$ LUTs are used to build a tracker.
Similarly, less than $0.05\%$ of flip-flops and less than $0.31\%$ LUTs are used to build a control unit.
If applying DFT-FPGA on large scale FPGAs, the utilization ratio will be further reduced.
The maximum tracking cycles are $Max$ $Cycle$ = $t*count_{max}$ which is illustrated in in section~\ref{trackers}. For example, 8-bit tracker can count a function with up to $65025=255*255$ cycles.

The control unit keeps SLICE addresses and indexes it after reading trackers' status. The control unit's resource utilization is shown in Table~\ref{curesource}. In this table, we show the resource of a control unit with one tracker when tracker size scales from $4bit$ to $9bit$.
A $n$ $bit$ counter needs $count_{max}$ depth in BRAM to store SLICE address because the $count$ start over when enter each iteration of $t$.
Thus, the depth needed for a tracker is much smaller than the length of a function.
When the counter is less than $8bit$,
Vivado HLS optimizes it into flip-flops and LUTs to save BRAM resource.

The timing performance of a DFT-FPGA framework is shown in Table~\ref{frequency}.
In this evaluation, we measure the maximum working frequency of 8-bit standalone tracker, standalone control unit, and DFT-FPGA which is an integration of one 8-bit tracker and control unit.
We can observe that a standalone tracker can work up to $500MHz$ and a standalone control unit can work up to $400MHz$ on FPGA $xc7z020clg484$.
 With the evaluated NV-FF FPGA performance and the DFT-FPGA performance, we can conclude that the DFT-FPGA consumes a small amount of FPGA resources and will not be the speed bottleneck after being applied to the source program.

\begin{table}[htbp]
\caption{Tracker Resource Utilization}
\label{Trackerresource}
\footnotesize
\begin{tabular}{cccccccc}
\hline
Tracker size & 4bit & 5bit & 6bit & 7bit  & 8bit  & 9bit   & FPGA   \\ \hline
FFs          & 15   & 18   & 21   & 24    & 27    & 30(0.03\%)     & 106400 \\
LUTs         & 90   & 102  & 102  & 102   & 102   & 110(0.21\%)    & 53200  \\
Max Cycle    & 225  & 961  & 3969 & 16129 & 65025 & 261121 & -      \\ \hline
\end{tabular}
\end{table}

\begin{table}[htbp]
\caption{Control Unit Resource Utilization}
\label{curesource}
\footnotesize
\begin{tabular}{cccccccc}
\hline
Tracker size & 4bit   & 5bit   & 6bit   & 7bit   & 8bit & 9bit & FPGA   \\ \hline
FFs          & 52     & 52     & 52     & 52(0.05\%)     & 36   & 36   & 106400 \\
LUTs         & 138    & 142    & 150    & 166(0.31\%)    & 134  & 134  & 53200  \\
BRAM         & $\sim$ & $\sim$ & $\sim$ & $\sim$ & 2    & 2    & 280      \\ \hline
\end{tabular}
\end{table}

\begin{table}[htbp]
\caption{DFT-FPGA Working Frequency}
\label{frequency}
\footnotesize
\begin{tabular}{cccccccc}
\hline
Frequency    & 100MHz  & 150MHz  & 200MHz  & 400MHz  & 450MHz   & 500MHz   & 600MHz   \\ \hline
Tracker      & $\surd$ & $\surd$ & $\surd$ & $\surd$ & $\surd$  & $\surd$  & $\times$ \\
Control Unit & $\surd$ & $\surd$ & $\surd$ & $\surd$ & $\times$ & $\times$ & $\times$ \\
DFT-FPGA   & $\surd$ & $\surd$ & $\surd$ & $\surd$ & $\times$ & $\times$ & $\times$ \\ \hline
\end{tabular}
\end{table}

\subsection{DFT-FPGA Case study}
\label{dftfpgaperformance}
In the previous section, we showed the performance of a tracker, a control unit and an individual DFT-FPGA framework.
In this section, we evaluate the performance of HLS benchmarks from $CHStone$~\cite{hara2009proposal} on FPGA $xc7z020clg484$ with/without applying DFT-FPGA under 400 $MHz$ working frequency.
The benchmarks cover large size benchmark such as $adpcm$ and small size benchmark such as $struct$.
Table~\ref{p1} and ~\ref{p2} show the resource utilization of source benchmarks and benchmarks applyied DFT-FPGA (DFT). In the table, resources such as BRAM, DSP, flip-flop, LUT, and SLICE are listed.
For the source benchmarks, the resource usage and utilization level are presented. 
For DFT-FPGA, its resource usage includes source benchmark and the applied DFT-FPGA framework.
The resource usage and utilization level increment are shown in the table for DFT-FPGA.
% From the evaluated benchmarks, we can observe that limited resource are occupied by DFT-FPGA components and DFT-FPGA shows good adaptability to different benchmarks.
For benchmarks with higher resource utilization, such as $adpcm$, $aes$, $gsm$, and $float$, a less than $1.5\%$ resource increment is observed.
DFT-FPGA generates trackers and control unit to track their data flow.
For small benchmarks such as $global$ and $struct$, less than $0.3\%$ resource increment is observed. 
Such benchmarks consume a small number of flip-flops and LUTs, which is even less than its tracker's resource requirement.
For such benchmarks, DFT-FPGA only generates control unit to keep all associated SLICE addresses and  save all data.
In this way, DFT-FPGA achieves resource efficiency in different sizes of benchmarks.
The proposed design shows good adaptability when source programs' size scales up and scales down.
It consumes a small number of FPGA resources to achieve intermediate data tracking in the analyzed benchmarks. 

\begin{table}[htbp]
\caption{DFT-FPGA in Different Benchmarks}
\label{p1}
\footnotesize
\begin{tabular}{ccc|cc|cc|c}
\hline
Benchmarks & \multicolumn{2}{c|}{adpcm}    & \multicolumn{2}{c|}{aes}      & \multicolumn{2}{c|}{gsm}     & FPGA   \\ \hline
           & source       & DFT            & source       & DFT            & source      & DFT            &        \\ \hline
BRAM       & 14(5\%)      & 16 (+0.7\%)    & 9(3\%)       & 11 (+0.07\%)   & 7(2.5\%)    & 9 (+0.7\%)     & 280    \\
DSP        & 114(51.8\%)  & 114(+0\%)      & 0(0\%)       & 0(+0\%)        & 46(20.9\%)  & 46(+0\%)       & 220    \\
FF         & 3464(3.3\%)  & 3769 (+0.28\%) & 1028(0.96\%) & 1321 (+0.28\%) & 1435(1.3\%) & 1682 (+0.23\%) & 106400 \\
LUT        & 8293(15.6\%) & 8928 (+1.2\%)  & 4552(8.6\%)  & 5330 (+1.5\%)  & 3903(7.3\%) & 4487 (+1\%)    & 53200  \\
SLICE      & -            & 472            & -            & 166            & -           & 211            & 13300  \\ \hline
\end{tabular}
\end{table}

\begin{table}[htbp]
\caption{DFT-FPGA in Different Benchmarks}
\label{p2}
\footnotesize
\begin{tabular}{ccc|cc|cc|c}
\hline
Benchmarks & \multicolumn{2}{c|}{float}   & \multicolumn{2}{c|}{global} & \multicolumn{2}{c|}{struct} & FPGA   \\ \hline
           & source      & DFT            & source      & DFT           & source      & DFT           &        \\ \hline
BRAM       & 14(5\%)     & 16 (+0.7\%)    & 3(1\%)      & 3 (+0\%)      & 4(1.4\%)    & 4(+0\%)       & 280    \\
DSP        & 14(6.4\%)   & 14(+0\%)       & 0(0\%)      & 0(+0\%)       & 0(0\%)      & 0(+0\%)       & 220    \\
FF         & 2997(2.8\%) & 3210 (+0.2\%)  & 74(0.07\%)  & 284 (+0.19\%) & 58(0.05\%)  & 267 (+0.19\%) & 106400 \\
LUT        & 4901(9.2\%) & 5150 (+0.47\%) & 713(1.3\%)  & 882 (+0.3\%)  & 228(0.4\%)  & 380 (+0.28\%) & 53200  \\
SLICE      & -           & 402            & -           & 36            & -           & 34            & 13300  \\ \hline
\end{tabular}
\end{table}

\subsection{Comparison DFT-FPGA with Checkpoint Technology}
\label{compare}
In the previous section, we evaluated the resource utilization of DFT-FPGA in different benchmarks.
In this section, we will present the comparison between DFT-FPGA and periodical checkpoint CP-FPGA in resource, roll-back time, and flip-flop storing under different power conditions.

As CP-FPGA places checkpoints at the end of all states and saves checkpoint to a BRAM, the number of extra BRAMs brought by CP-FPGA equals the number of states in a benchmark.
% Though a state can be called multiple times and checkpoint is placed after each call.
In the proposed design, the extra BRAM is brought by control unit, which stores SLICE addresses.
In table~\ref{bramcompare}, we show the number of states in each benchmark and BRAM utilization in DFT-FPGA and CP-FPGA.
In FPGA $xc7z020clg484$, a single BRAM stores $18Kb$ data.
As shown in the table, the BRAM resource is significantly decreased in DFT-FPGA.
This is because the BRAM in DFT-FPGA is generated according to the number of SLICEs needed to be tracked.
Each BRAM in DFT-FPGA is fully utilized.
BRAM in CP-FPGA can not be fully utilized as it is assigned by state number, regardless of the size of state result.

\begin{table}[ht]
\caption{BRAM usage in DFT-FPGA and CP-FPGA}
\label{bramcompare}
\footnotesize
\begin{tabular}{ccccccc}
\hline
Benchmarks & adpcm   & aes   & gsm      & float & global & struct \\ \hline
states     & 24      & 8     & 13       & 1     & 3      & 2      \\
BRAM\_DFT  & 2       & 2     & 2        & 2     & 0      & 0      \\
BRAM\_CP   & 24(12x) & 8(4x) & 13(6.5x) & 1     & 3(3x)  & 2(2x)  \\ \hline
\end{tabular}
\end{table}

\begin{figure}[htbp]
\centering
\begin{minipage}[t]{0.48\textwidth}
\centering
\includegraphics[width=6cm]{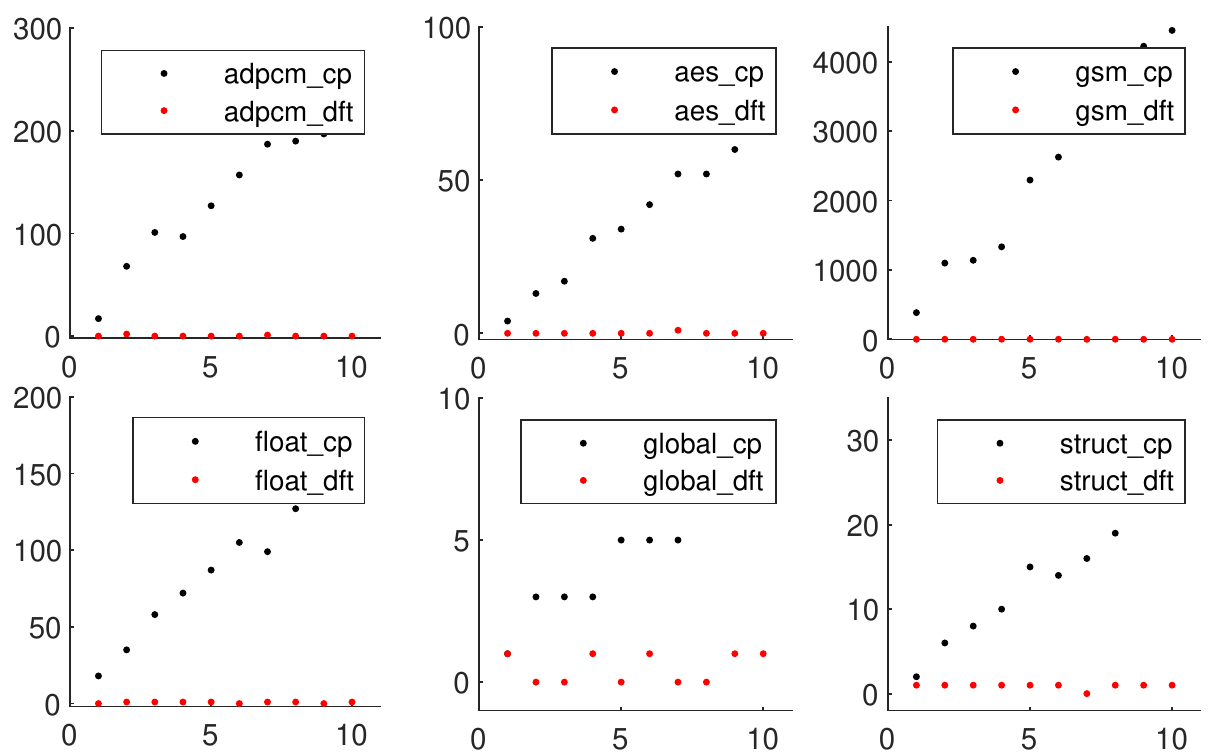}
\caption{Roll-back time in CP-FPGA and DFT-FPGA.}\label{expd1}
\end{minipage}
\begin{minipage}[t]{0.48\textwidth}
\centering
\includegraphics[width=6cm]{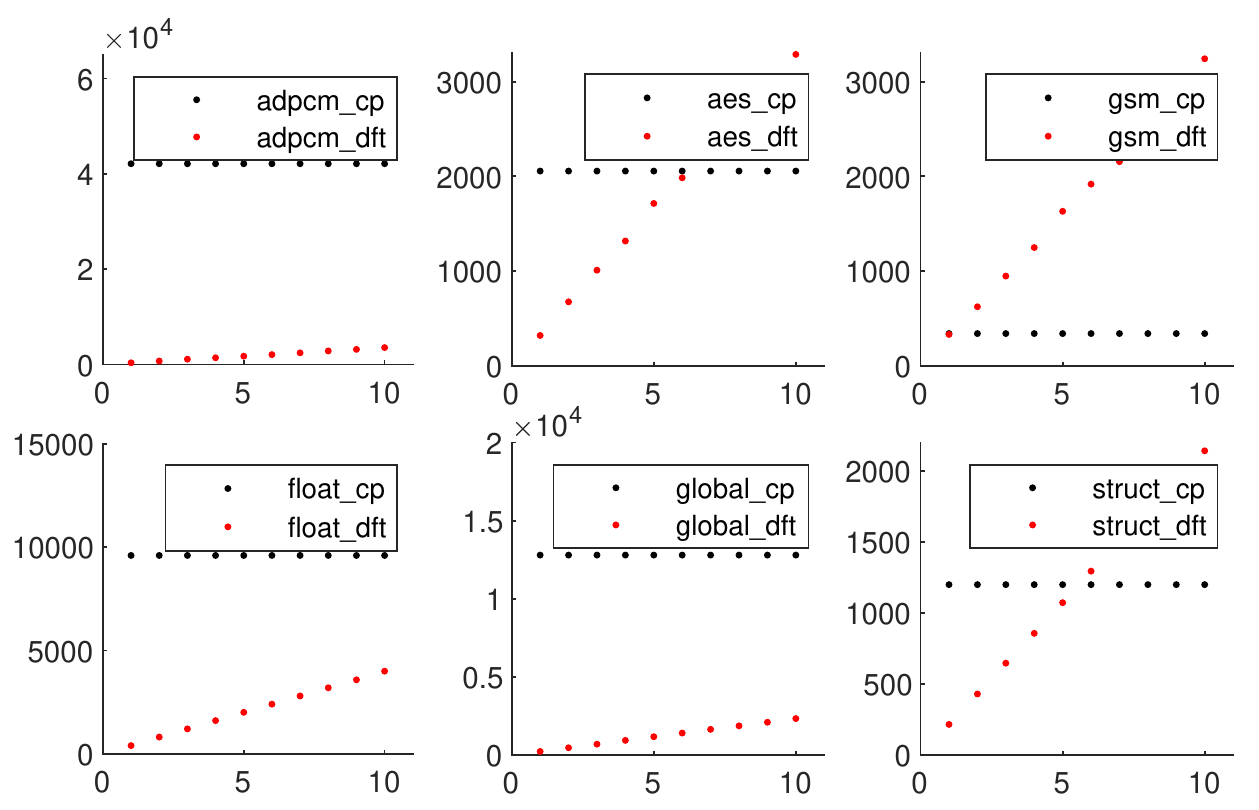}
\caption{Flip-flop operations in CP-FPGA and DFT-FPGA.}\label{expd2}
\end{minipage}
\end{figure}

The roll-back time and the number of stored flip-flops in different benchmarks are shown in Figure~\ref{expd1} and Figure~\ref{expd2}.
In Fig.~\ref{expd1}, the x-axis shows the number of power lost during benchmarks running and the y-axis shows the roll-back clock cycles.
Power lost is randomly triggered within a benchmark's computation length, e.g at power lost is 5, there are 5 power lost during computing and every power lost is randomly triggered.
In the evaluation, we simulate the number of power lost from 1 to 10 during the computation to mimic different power conditions.
Without loss of generality, for each power lost case, we record the result of its mean of 10 test rounds.
As shown in the figure, the roll-back time for the proposed design is near zero in all benchmarks and all power conditions.
Our performance is not influenced even if power condition is worse. 
This is because the data flow inside each state is aware by DFT-FPGA;
DFT-FPGA can retrieve computation from where it is interrupted.
For CP-FPGA, it needs to find its nearest checkpoint and recover from that point.
If the length of states is long in benchmarks, the interval between two checkpoints is far from each other. It causes long roll-back if power is lost during the middle of such state.
This is observed in the $gsm$ benchmark which consists of multiple states with over one thousand cycles each.
When power lost occurs in one of such states, long roll-back happens.
The performance of periodical placing checkpoint technology is significantly influenced by the source program and power condition.
Meanwhile, the proposed DFT-FPGA shows good adaptability in minimized roll-back time for different benchmarks in different power conditions.

Figure~\ref{expd2} shows the stored flip-flop data for benchmarks in different power conditions. In this figure, we record the number of flip-flops that are stored in CP-FPGA and the proposed design.
The x-axis shows the power lost case ranging from 1 to 10. The y-axis shows the number of flip-flops that are stored after finishing computation.
As CP-FPGA periodically places checkpoints in design, the number of flip-flops stored is a constant in all power conditions.
After a state is called, it stores its state result data in registers to BRAM as a checkpoint.
It leads to unnecessary data storing when power lost happens occasionally such as once or twice in computation.
For DFT-FPGA, the number of flip-flops which is saved consists of both intermediate data registers and trackers' flip-flop resources.
We can observe linear increasing in flip-flops when the number of power lost becomes worse.
This is because the data storage in DFT-FPGA happens only at power lost.
Periodical placing checkpoint may have fewer
flip-flops usages if a small number of checkpoints are placed.
Such as $truct$, it is arranged to several long states.
CP-FPGA have less flip-flop storing in such benchmarks. However, it brings long roll-back time as shown in Fig.~\ref{expd1}.
% The proposed design outperforms periodical checkpoint technology in both roll-back time and flip-flop storage. 

% \begin{table}[htbp]
% \caption{BRAM Usage in DFT-FPGA and CP-FPGA}
% \label{bramcompare}
% \footnotesize
% \begin{tabular}{ccccccc}
% \hline
% Benchmarks & adpcm   & aes   & gsm      & float & global & struct \\ \hline
% states     & 24      & 8     & 13       & 1     & 3      & 2      \\
% BRAM\_DFT  & 2       & 2     & 2        & 2     & 0      & 0      \\
% BRAM\_CP   & 24(12x) & 8(4x) & 13(6.5x) & 1     & 3(3x)  & 2(2x)  \\ \hline
% \end{tabular}
% \end{table}

\section{Conclusion}

We propose a data-flow tracking framework, DFT-FPGA, for non-volatile FPGA. It is a full High-Level-Synthesis based framework targeting non-volatile FPGA that can online track and locate the physical location of intermediate data. By parsing, storing, and retrieving certain area of FPGA SLICE, the proposed design can assist NV-FPGA in intermittent computing with minimum resource overhead.
The proposed DFT-FPGA also shows good adaptability in different benchmarks under various power conditions with better resource utilization and less roll-back time. 

\section{Acknowledgements}
This work was supported in part by the National
Science Foundation under Grant CCF-1820537, in part by the National Natural
Science Foundation of China under Grant 61934005, in part by the National Natural
Science Foundation of China under Grant 61674094, in part by the National Science Foundation under Grant CCF-1527464, and in part by the Research Grants Council
of the Hong Kong Special Administrative Region, China under Grant CityU 11278316.

%
% The next two lines define the bibliography style to be used, and the bibliography file.
\bibliographystyle{unsrt}
\bibliography{Reference}

\begin{thebibliography}{10}

\bibitem{xu2018scaling}
Xiaowei Xu, Yukun Ding, Sharon~Xiaobo Hu, Michael Niemier, Jason Cong, Yu~Hu,
  and Yiyu Shi.
\newblock Scaling for edge inference of deep neural networks.
\newblock {\em Nature Electronics}, 1(4):216, 2018.

\bibitem{xu2017edge}
Xiaowei Xu, Qing Lu, Tianchen Wang, Jinglan Liu, Cheng Zhuo, Xiaobo~Sharon Hu,
  and Yiyu Shi.
\newblock Edge segmentation: Empowering mobile telemedicine with compressed
  cellular neural networks.
\newblock In {\em Proceedings of the 36th International Conference on
  Computer-Aided Design}, pages 880--887. IEEE Press, 2017.

\bibitem{xu2017efficient}
Xiaowei Xu, Dewen Zeng, Wenyao Xu, Yiyu Shi, and Yu~Hu.
\newblock An efficient memristor-based distance accelerator for time series
  data mining on data centers.
\newblock In {\em 2017 54th ACM/EDAC/IEEE Design Automation Conference (DAC)},
  pages 1--6. IEEE, 2017.

\bibitem{lattice}
Lattice.
\newblock ice40 lp/hx/lm.
\newblock {\em lattice}, 2017.

\bibitem{Tang2014}
Xifan Tang, Pierre-Emmanuel Gaillardon, and Giovanni De~Micheli.
\newblock A high-performance low-power near-vt rram-based fpga.
\newblock In {\em 2014 International Conference on Field-Programmable
  Technology (FPT)}, pages 207--214. IEEE, 2014.

\bibitem{Tang2016}
Xifan Tang, Gain Kim, Pierre-Emmanuel Gaillardon, and Giovanni De~Micheli.
\newblock A study on the programming structures for rram-based fpga
  architectures.
\newblock {\em IEEE Transactions on Circuits and Systems I: Regular Papers},
  63(4):503--516, 2016.

\bibitem{zhao2009spin}
Weisheng Zhao, Eric Belhaire, Claude Chappert, and Pascale Mazoyer.
\newblock Spin transfer torque (stt)-mram--based runtime reconfiguration fpga
  circuit.
\newblock {\em ACM Transactions on Embedded Computing Systems (TECS)}, 9(2):14,
  2009.

\bibitem{guo2010resistive}
Xiaochen Guo, Engin Ipek, and Tolga Soyata.
\newblock Resistive computation: avoiding the power wall with low-leakage,
  stt-mram based computing.
\newblock In {\em ACM SIGARCH Computer Architecture News}, volume~38, pages
  371--382. ACM, 2010.

\bibitem{gaillardon2010}
Pierre-Emmanuel Gaillardon, M~Haykel Ben-Jamaa, Marina Reyboz, Giovanni~Betti
  Beneventi, Fabien Clermidy, Luca Perniola, and Ian O'Connor.
\newblock Phase-change-memory-based storage elements for configurable logic.
\newblock In {\em 2010 International Conference on Field-Programmable
  Technology (FPT)}, pages 17--20. IEEE, 2010.

\bibitem{Chen2010}
Yibo Chen, Jishen Zhao, and Yuan Xie.
\newblock 3d-nonfar: three-dimensional non-volatile fpga architecture using
  phase change memory.
\newblock In {\em Proceedings of the 16th ACM/IEEE international symposium on
  Low power electronics and design}, pages 55--60. ACM, 2010.

\bibitem{lee2017reram}
Albert Lee, Chieh-Pu Lo, Chien-Chen Lin, Wei-Hao Chen, Kuo-Hsiang Hsu, Zhibo
  Wang, Fang Su, Zhe Yuan, Qi~Wei, Ya-Chin King, and Chrong-Jung Lin.
\newblock A reram-based nonvolatile flip-flop with self-write-termination
  scheme for frequent-off fast-wake-up nonvolatile processors.
\newblock {\em IEEE Journal of Solid-State Circuits}, 52(8):2194--2207, 2017.

\bibitem{patil2019fpga}
Vilabha~S Patil, Yashwant~B Mane, and Shraddha Deshpande.
\newblock Fpga based power saving technique for sensor node in wireless sensor
  network (wsn).
\newblock In {\em Computational Intelligence in Sensor Networks}, pages
  385--404. Springer, 2019.

\bibitem{bengherbia2017fpga}
Billel Bengherbia, Mohamed~Ould Zmirli, Abdelmoghni Toubal, and Abderrezak
  Guessoum.
\newblock Fpga-based wireless sensor nodes for vibration monitoring system and
  fault diagnosis.
\newblock {\em Measurement}, 101:81--92, 2017.

\bibitem{obeid2016towards}
Abdulfattah~M Obeid, Fatma Karray, Mohamed~Wassim Jmal, Mohamed Abid,
  Syed~Manzoor Qasim, and Mohammed~S BenSaleh.
\newblock Towards realisation of wireless sensor network-based water pipeline
  monitoring systems: a comprehensive review of techniques and platforms.
\newblock {\em IET science, measurement \& technology}, 10(5):420--426, 2016.

\bibitem{xue2015analysis}
T~Xue and S~Roundy.
\newblock Analysis of magnetic plucking configurations for frequency
  up-converting harvesters.
\newblock In {\em Journal of Physics: Conference Series}, volume 660, page
  012098. IOP Publishing, 2015.

\bibitem{ma2017incidental}
Kaisheng Ma, Xueqing Li, Jinyang Li, Yongpan Liu, Yuan Xie, Jack Sampson,
  Mahmut~Taylan Kandemir, and Vijaykrishnan Narayanan.
\newblock Incidental computing on iot nonvolatile processors.
\newblock In {\em Proceedings of the 50th Annual IEEE/ACM International
  Symposium on Microarchitecture}, pages 204--218. ACM, 2017.

\bibitem{ma2015nonvolatile}
Kaisheng Ma, Xueqing Li, Shuangchen Li, Yongpan Liu, John~Jack Sampson, Yuan
  Xie, and Vijaykrishnan Narayanan.
\newblock Nonvolatile processor architecture exploration for energy-harvesting
  applications.
\newblock {\em IEEE Micro}, 35(5):32--40, 2015.

\bibitem{maeng2017alpaca}
Kiwan Maeng, Alexei Colin, and Brandon Lucia.
\newblock Alpaca: intermittent execution without checkpoints.
\newblock {\em Proceedings of the ACM on Programming Languages}, 1(OOPSLA):96,
  2017.

\bibitem{pan2017lightweight}
Chen Pan, Mimi Xie, Yongpan Liu, Yanzhi Wang, Chun~Jason Xue, Yuangang Wang,
  Yiran Chen, and Jingtong Hu.
\newblock A lightweight progress maximization scheduler for non-volatile
  processor under unstable energy harvesting.
\newblock In {\em Proceedings of the 18th ACM SIGPLAN/SIGBED Conference on
  Languages, Compilers, and Tools for Embedded Systems}, pages 101--110. ACM,
  2017.

\bibitem{balsamo2015hibernus}
Domenico Balsamo, Alex~S Weddell, Geoff~V Merrett, Bashir~M Al-Hashimi, Davide
  Brunelli, and Luca Benini.
\newblock Hibernus: Sustaining computation during intermittent supply for
  energy-harvesting systems.
\newblock {\em IEEE Embedded Systems Letters}, 7(1):15--18, 2015.

\bibitem{xie2015fixing}
Mimi Xie, Mengying Zhao, Chen Pan, Jingtong Hu, Yongpan Liu, and Chun~Jason
  Xue.
\newblock Fixing the broken time machine: Consistency-aware checkpointing for
  energy harvesting powered non-volatile processor.
\newblock In {\em Proceedings of the 52nd Annual Design Automation Conference},
  page 184. ACM, 2015.

\bibitem{xie2016checkpoint}
Mimi Xie, Mengying Zhao, Chen Pan, Hehe Li, Yongpan Liu, Youtao Zhang,
  Chun~Jason Xue, and Jingtong Hu.
\newblock Checkpoint aware hybrid cache architecture for nv processor in energy
  harvesting powered systems.
\newblock In {\em Proceedings of the Eleventh IEEE/ACM/IFIP International
  Conference on Hardware/Software Codesign and System Synthesis}, page~22. ACM,
  2016.

\bibitem{Mirhoseini}
Azalia Mirhoseini, Bita~Darvish Rouhani, Ebrahim Songhori, and Farinaz
  Koushanfar.
\newblock Chime: Checkpointing long computations on interm ittently energized
  iot devices.
\newblock {\em IEEE Transactions on Multi-Scale Computing Systems},
  2(4):277--290, 2016.

\bibitem{yuan2016cp}
Zhe Yuan, Yongpan Liu, Hehe Li, and Huazhong Yang.
\newblock Cp-fpga: Computation data-aware software/hardware co-design for
  nonvolatile fpgas based on checkpointing techniques.
\newblock In {\em Design Automation Conference (ASP-DAC), 2016 21st Asia and
  South Pacific}, pages 569--574. IEEE, 2016.

\bibitem{gaillardon2013design}
P-E Gaillardon, Davide Sacchetto, Giovanni~Betti Beneventi, M~Haykel~Ben Jamaa,
  Luca Perniola, Fabien Clermidy, Ian O'Connor, and Giovanni De~Micheli.
\newblock Design and architectural assessment of 3-d resistive memory
  technologies in fpgas.
\newblock {\em IEEE Transactions on Nanotechnology}, 12(1):40--50, 2013.

\bibitem{Cong2011}
Jason Cong and Bingjun Xiao.
\newblock mrfpga: A novel fpga architecture with memristor-based
  reconfiguration.
\newblock In {\em 2011 IEEE/ACM International Symposium on Nanoscale
  Architectures}, pages 1--8. IEEE, 2011.

\bibitem{zand2017radiation}
Ramtin Zand and Ronald~F DeMara.
\newblock Radiation-hardened mram-based lut for non-volatile fpga soft error
  mitigation with multi-node upset tolerance.
\newblock {\em Journal of Physics D: Applied Physics}, 50(50):505002, 2017.

\bibitem{jo2016variation}
Kangwook Jo, Kyungseon Cho, and Hongil Yoon.
\newblock Variation-tolerant and low power look-up table (lut) using
  spin-torque transfer magnetic ram for non-volatile field programmable gate
  array (fpga).
\newblock In {\em 2016 International SoC Design Conference (ISOCC)}, pages
  101--102. IEEE, 2016.

\bibitem{almurib2016design}
Haider Abbas~F Almurib, Thulasiraman~Nandha Kumar, and Fabrizio Lombardi.
\newblock Design and evaluation of a memristor-based look-up table for
  non-volatile field programmable gate arrays.
\newblock {\em IET Circuits, Devices \& Systems}, 10(4):292--300, 2016.

\bibitem{rajaei2016radiation}
Ramin Rajaei.
\newblock Radiation-hardened design of nonvolatile mram-based fpga.
\newblock {\em IEEE Transactions on Magnetics}, 52(10):1--10, 2016.

\bibitem{ju2018nvm}
Lei Ju, Xiaojin Sui, Shiqing Li, Mengying Zhao, Chun~Jason Xue, Jingtong Hu,
  and Zhiping Jia.
\newblock Nvm-based fpga block ram with adaptive slc-mlc conversion.
\newblock {\em IEEE Transactions on Computer-Aided Design of Integrated
  Circuits and Systems}, 37(11):2661--2672, 2018.

\bibitem{Wong2012}
H-S~Philip Wong, Heng-Yuan Lee, Shimeng Yu, Yu-Sheng Chen, Yi~Wu, Pang-Shiu
  Chen, Byoungil Lee, Frederick~T Chen, and Ming-Jinn Tsai.
\newblock Metal--oxide rram.
\newblock {\em Proceedings of the IEEE}, 100(6):1951--1970, 2012.

\bibitem{onkaraiah2012bipolar}
Santhosh Onkaraiah, Marina Reyboz, Fabien Clermidy, Jean-Michel Portal, Marc
  Bocquet, Chritophe Muller, Hraziia, Costin Anghel, and Amara Amara.
\newblock Bipolar reram based non-volatile flip-flops for low-power
  architectures.
\newblock In {\em 2012 IEEE 10th International New Circuits and Systems
  Conference (NEWCAS)}, pages 417--420. IEEE, 2012.

\bibitem{Jabeur2014}
K~Jabeur, G~Di~Pendina, and G~Prenat.
\newblock Ultra-energy-efficient cmos/magnetic nonvolatile flip-flop based on
  spin-orbit torque device.
\newblock {\em Electronics Letters}, 50(8):585--587, 2014.

\bibitem{Chien2016}
Tsai-Kan Chien, Lih-Yih Chiou, Yao-Chun Chuang, Shyh-Shyuan Sheu, Heng-Yuan Li,
  Pei-Hua Wang, Tzu-Kun Ku, Ming-Jinn Tsai, and Chih-I~Wu Wu.
\newblock A low store energy and robust reram-based flip-flop for normally off
  microprocessors.
\newblock In {\em 2016 IEEE International Symposium on Circuits and Systems
  (ISCAS)}, pages 2803--2806. IEEE, 2016.

\bibitem{floor}
Xilinx.
\newblock Xilinx floorplanning methodology guide.
\newblock {\em Xilinx}, 2009.

\bibitem{nane2015survey}
Razvan Nane, Vlad-Mihai Sima, Christian Pilato, Jongsok Choi, Blair Fort,
  Andrew Canis, Yu~Ting Chen, Hsuan Hsiao, Stephen Brown, Fabrizio Ferrandi,
  et~al.
\newblock A survey and evaluation of fpga high-level synthesis tools.
\newblock {\em IEEE Transactions on Computer-Aided Design of Integrated
  Circuits and Systems}, 35(10):1591--1604, 2015.

\bibitem{navarro2013high}
Denis Navarro, {\'O}scar Luc{\i}, Luis~A Barrag{\'a}n, Isidro Urriza, Oscar
  Jimenez, et~al.
\newblock High-level synthesis for accelerating the fpga implementation of
  computationally demanding control algorithms for power converters.
\newblock {\em IEEE Transactions on Industrial Informatics}, 9(3):1371--1379,
  2013.

\bibitem{jiang2019achieving}
Weiwen Jiang, Edwin H-M Sha, Xinyi Zhang, Lei Yang, Qingfeng Zhuge, Yiyu Shi,
  and Jingtong Hu.
\newblock Achieving super-linear speedup across multi-fpga for real-time dnn
  inference.
\newblock {\em ACM Transactions on Embedded Computing Systems (TECS)},
  18(5s):67, 2019.

\bibitem{jiang2019accuracy}
Weiwen Jiang, Xinyi Zhang, Edwin H-M Sha, Lei Yang, Qingfeng Zhuge, Yiyu Shi,
  and Jingtong Hu.
\newblock Accuracy vs. efficiency: Achieving both through fpga-implementation
  aware neural architecture search.
\newblock {\em arXiv preprint arXiv:1901.11211}, 2019.

\bibitem{Canis2013}
Andrew Canis, Jongsok Choi, Mark Aldham, Victor Zhang, Ahmed Kammoona, Tomasz
  Czajkowski, Stephen~D Brown, and Jason~H Anderson.
\newblock Legup: An open-source high-level synthesis tool for fpga-based
  processor/accelerator systems.
\newblock {\em ACM Transactions on Embedded Computing Systems (TECS)},
  13(2):24, 2013.

\bibitem{ahmed2018towards}
Saad Ahmed, Muhammad~Hamad Alizai, Junaid~Haroon Siddiqui, Naveed~Anwar Bhatti,
  and Luca Mottola.
\newblock Towards smaller checkpoints for better intermittent computing.
\newblock In {\em 2018 17th ACM/IEEE International Conference on Information
  Processing in Sensor Networks (IPSN)}, pages 132--133. IEEE, 2018.

\bibitem{zhang2018low}
Xinyi Zhang, Clay Patterson, Yongpan Liu, Chengmo Yang, Chun~Jason Xue, and
  Jingtiong Hu.
\newblock Low overhead online checkpoint for intermittently powered
  non-volatile fpgas.
\newblock In {\em 2018 IEEE Computer Society Annual Symposium on VLSI
  (ISVLSI)}, pages 238--244. IEEE, 2018.

\bibitem{49dff}
Masood Qazi, Amerasekera Ajith, and Anantha~P. Chandrakasan.
\newblock A 3.4pj feram-enabled d flip-flop in 0.13µm cmos for nonvolatile
  processing in digital systems.
\newblock In {\em 2013 IEEE International Solid-State Circuits Conference
  Digest of Technical Papers}. IEEE, 2013.

\bibitem{khan2017study}
Farheen~Fatima Khan and Andy Ye.
\newblock A study on the accuracy of minimum width transistor area in
  estimating fpga layout area.
\newblock {\em Microprocessors and Microsystems}, 52:287--298, 2017.

\bibitem{cadambi2010programmable}
Srihari Cadambi, Abhinandan Majumdar, Michela Becchi, Srimat Chakradhar, and
  Hans~Peter Graf.
\newblock A programmable parallel accelerator for learning and classification.
\newblock In {\em Proceedings of the 19th international conference on Parallel
  architectures and compilation techniques}, pages 273--284. ACM, 2010.

\bibitem{chakradhar2010dynamically}
Srimat Chakradhar, Murugan Sankaradas, Venkata Jakkula, and Srihari Cadambi.
\newblock A dynamically configurable coprocessor for convolutional neural
  networks.
\newblock {\em ACM SIGARCH Computer Architecture News}, 38(3):247--257, 2010.

\bibitem{hara2009proposal}
Yuko Hara, Hiroyuki Tomiyama, Shinya Honda, and Hiroaki Takada.
\newblock Proposal and quantitative analysis of the chstone benchmark program
  suite for practical c-based high-level synthesis.
\newblock {\em Journal of Information Processing}, 17:242--254, 2009.

\end{thebibliography}

\end{document}